\documentclass{aa}

\usepackage[utf8]{inputenc}
\usepackage{graphicx}
\usepackage{natbib}
\usepackage{array}
\usepackage{makecell}
\usepackage[table,xcdraw]{xcolor}
\usepackage{multirow}
\usepackage{amsmath}

\begin{document}

\title{Geometrical properties of the interaction between \\ oblique incoming coronal waves and coronal holes}

\titlerunning{Geometrical properties of CW-CH interaction}
\authorrunning{Piantschitsch, I., Terradas, J.}

\author{Piantschitsch, I.$^{1,2}$, Terradas, J.$^1$}

\institute{$^1$Departament de F\'\i sica, Universitat de les Illes Balears (UIB),
E-07122, Spain \\   Institute of Applied Computing \& Community Code (IAC$^3$),
UIB, Spain\\
$^2$Institute of Physics, University of Graz, Universit\"atsplatz 5, A-8010 Graz, Austria
\\
\email{isabell.piantschitsch@uib.es}
}

\date{\today}

\abstract{Observations of coronal waves (CWs) interacting with coronal holes (CHs) show the formation of typical wave-like features, such as reflected, refracted and transmitted waves (collectively, secondary waves). In accordance with these observations, numerical evidence for the wave characteristics of CWs is given by simulations which demonstrate effects of deflection and reflection when a CW interacts with regions exhibiting a sudden density drop, such as CHs. However, secondary waves are usually weak in their signal and simulations are limited in the way the according idealisations have to be chosen. Hence, several properties of the secondary waves during a CW-CH interaction are unclear or ambiguous and might lead to misinterpretations. In this study we follow a theoretical approach and focus in particular on the geometrical properties of secondary waves caused by the interaction between oblique incoming CWs and CHs. Based on a linear theory, we derive analytical expressions for reflection and transmission coefficients, which tell us how strongly the amplitudes of the secondary waves increase and decrease with respect to the incoming wave, respectively. Additionally, we provide analytical terms for crucial incidence angles that are capable of giving information about the energy flux, the phase and the reflection properties of the secondary waves. These novel expressions provide a supplementary tool for estimating CW properties in a fast and straightforward way and, therefore, might have relevant consequences for a possible new interpretation of already studied CW-CH interaction events and the clarification of ambiguous observational data.}


\maketitle

\section{Introduction}

Large-scale propagating disturbances in the corona were directly observed for the first time by the Extreme-ultraviolet Imaging Telescope (EIT; \citealt{Delaboudiniere1995}) onboard the Solar and Heliospheric Observatory \citep{Domingo1995}. These so-called EIT waves (often also referred to as EUV waves or coronal bright fronts) can be observed over the entire solar surface and are commonly associated with energetic eruptions such as coronal mass ejections (CMEs) or flares \citep[see e.g.][]{vrsnaklulic2000}. While CMEs are a necessary, though not sufficient, condition for the occurrence of EUV waves, their connection with solar flares is generally considered as weak, nonetheless, there is evidence of an energetic correlation between these three phenomena \citep[see e.g.][]{Wei2014}.

Within the last two decades, three main theories have evolved, trying to explain the nature of EUV waves by either using a wave model, a non-wave model or, alternatively, a hybrid model. Wave theories interpret EUV waves as fast-mode magnetohydrodynamic (MHD) waves \citep{thompson1998,vrsnaklulic2000,wang2000,wu2001,Warmuth2004,patsourakos2009,patsourakos_vourlidas2009, schmidt_ofman_2010,veronig2010,lulic2013}, whereas pseudo-wave theories consider them as the result of the reconfiguration of the coronal magnetic field, caused by either Joule heating \citep{delannee_aulanier1999, delannee2000}, continuous small-scale reconnection \citep{Attrill2007a,Attrill2007b}, or stretching of magnetic field lines \citep{Chen2002}. Hybrid models, on the other hand, try to explain the observed disturbances in the corona by combining the two competing approaches of wave and non-wave model. In this approach the outer envelope of a CME is interpreted as a reconfiguration of the magnetic field, sometimes also named "pseudo-wave", that is followed by a freely propagating fast-mode MHD wave  \citep{Chen2002,Chen2005,Zhukov_Auchere_2004, Cohen2009, Liu2010, Chen_Wu_2011, Downs2011,Cheng2012}.

In this study, we follow an approach in which EUV waves are considered to be true waves, particularly focusing on the geometrical properties these waves exhibit when interacting with a low-density region such as a coronal hole (CH). For the reason of simplicity and due to the fact that other authors which also support the wave model use a similar terminology, we will hereafter refer to fast-mode MHD waves in the corona as coronal waves (CWs).

The interaction of CWs with CHs results, among other effects, in the formation of reflected, refracted and transmitted waves, which confirm their interpretation as fast mode MHD waves \citep{vrsnaklulic2000,wang2000,wu2001,Warmuth2004,veronig2010}. Observational evidence for their wave-like behaviour is given by authors who report waves being reflected and refracted at a CH boundary \citep[e.g.][]{ Long2008, gopal2009,kienreichetal2013}, transmitted through a CH \citep{liu19,olmedoetal2012} or partially penetrated into a CH \citep[e.g.][]{Veronig2006}. These observational findings are confirmed by numerical simulations describing effects such as deflection, reflection and transmission when a CW interacts with a low density region such as a CH \citep{Piantschitsch2017,afanasyev2018,Piantschitsch2018a, Piantschitsch2018b}.

From observations of CW-CH interactions we obtain measurements of different CW parameters, such as the density and the velocity amplitudes as well as the phase speed of secondary waves \citep[e.g.][]{Muhr2011,Kienreich2011}. However, due to their weak signals, measurements of secondary waves have so far been provided with rather low quality and accuracy. In particular, the density distribution inside of a CH can hardly be obtained in general since the density contrast to its surrounding is large and only few studies provide statistical information about these parameters \citep[e.g.][]{heinemann2019,Saqri2020}.

Studying CW-CH interaction and its resulting secondary waves is important for several reasons. Among them are, for example, observations of reflected waves which have been reported to have much higher phase speeds than the incoming wave for which there is, as yet, no conclusive explanation \citep{gopal2009,podladchikova2019}. In addition, CWs and their interaction with CHs provide a lot of information about CHs themselves, particularly about their boundaries. This is of great importance since a change in the shape of the CH and, consequently, a change in the position of the coronal hole boundary (CHB) influences predictions of high-speed solar wind streams \citep{riley2015}. Another reason for investigating CW-CH interaction are latest observations which use EUV waves as a diagnostic tool to infer the physical conditions of the solar corona on a global scale \citep{liu19}. The wave propagation in this study resulted in strong reflections and transmission in and out of both polar coronal holes. This observed event and its remarkable wave characteristics have the potential to allow global coronal seismology which was first introduced by \citet{West2011} and \citet{Long2013}, and which is an area yet to be fully exploited.

In general there seems to be a lack of theoretical studies about CW-CH interaction, therefore, a first and simple theoretical approach has been developed in \citet{Piantschitsch2020} in which analytical expressions for CW parameters in the case of a perpendicular incoming CW with respect to the CH have been derived. These theoretical estimations provide a fast and straightforward method to calculate CW and CH parameters by using easily available and basic observational measurements. 

However, the perpendicular case is valid only for a small range of interaction events and needs to be extended to an approach including oblique incoming CWs with respect to the CHB. The aim of this paper is therefore the continuation and the extension of the theoretical studies about CW-CH interaction, and in particular, studying and analysing the influence of the incidence angle on the parameters of secondary waves, such as the angle of the transmitted waves as well as the resulting density and velocity amplitudes. Analogously to the perpendicular case, we will derive analytical expressions for the transmission and the reflection coefficients, which describe the changes of the amplitudes of the transmitted and the reflected waves with respect to the incoming wave. As in \citet{Piantschitsch2020}, the derivation of these terms is based on a linear theory which is in accordance to observations that report CWs to be only weakly nonlinear \citep[e.g.][]{Muhr2011}. Also, the comparison of the analytically derived expressions with results of simulations of weakly nonlinear MHD waves \citep[see][]{Piantschitsch2020} showed good agreement and encourages therefore the continuation of the linear approach used in this study.

In Sect.~\ref{model_equat} we describe the setup for our model and derive the analytical expressions for the reflection and the transmission coefficients, followed by a comprehensive description of the different incidence angles as well as an analysis of the related phase properties in Sect.~\ref{angles}. In Sect.~\ref{wave_flux} we analyse the properties of the wave energy flux perpendicular and parallel to the interface, respectively. In the subsequent Sects.~\ref{refl_wave} and \ref{transm_wave} we describe and analyse the density and the velocity fluctuations of the reflected and the transmitted waves, respectively. We discuss the conclusions that can be drawn from our theoretical results in Sect.~\ref{conclusions} and provide a link to an online tool that can be used to visualise the results presented in this paper.

\section{Model and reflection/transmission coefficients} \label{model_equat}

\subsection{Homogeneous medium}

We consider a simple equilibrium including a vertical and constant magnetic field, ${\bf B_0}=B_0\, {\bf \hat{e}}_z$, pointing in the $z-$direction, and a constant density distribution, $\rho_0$. We assume that the gas pressure is small compared to the magnetic field and will therefore be neglected. The linearised version of the nonlinear MHD equations is obtained by assuming that all variables are written as the sum of an equilibrium term (with subindex 0) and a linear term. The process of linearisation, retaining terms up to first order leads, for the perturbed variables, to the following standard continuity equation
\begin{equation}\label{cont0}
    \frac{\partial\rho}{\partial t}+\nabla \cdot (\rho_0 {\bf v})=0,
\end{equation}
and momentum equation
\begin{equation}\label{moment0}
    \rho_0 \frac{\partial {\bf v}}{\partial t}=- \nabla p_{\rm T},
\end{equation}
where $\rho_0$ is the equilibrium density and
\begin{equation}\label{ptot0}
    p_{\rm T}=\frac{\bf{B_0}\cdot{\bf b}}{\mu_0},
\end{equation}
is the total perturbed pressure that only contains the magnetic term. The linearised induction equation is
\begin{equation}\label{induction0}
    \frac{\partial {\bf b}}{\partial t}=\nabla\times\left({\bf v}\times\bf{B_0}\right).
\end{equation}
Since the system is invariant in the $z-$direction and there is no gas pressure we have that 
${\bf v}=\left(v_x,v_y,0\right)$ and 
${\bf b}=\left(0,0,b_z\right)$. The magnetic pressure perturbation is $p_{\rm T}=B_0\, b_z/\mu_0$.

The characteristic speed of the medium is the Alfv\'en velocity, $v_{\rm A}=B_0/\sqrt{\mu_0 \rho_0}$. For a homogeneous  equilibrium $v_{\rm A}$ is constant and fluctuations in the system propagate as simple plane waves. We assume that
\begin{equation}\label{simplewave}
    {\bf v}={\bar {\bf v}}\,e^{i\left(\omega t- k_x x-k_y y\right)},
\end{equation}
and the same for the perturbed magnetic field and the density fluctuation. 
By using the expressions from the previous linearised equations and performing the temporal and the spatial derivatives we obtain that the amplitudes of the perturbed variables (for simplicity we do not use the overbars hereafter) are
\begin{align}
    v_x&=\frac{1}{\rho_0}\frac{k_x}{\omega}p_{\rm T},\label{vx}\\
    v_y&=\frac{1}{\rho_0}\frac{k_y}{\omega}p_{\rm T},\label{vy}\\
    p_{\rm T}&=\frac{B^2_0}{\mu_0}\frac{1}{\omega}\left(k_x v_x+k_y v_y\right),\label{pt}\\
   \rho &= \rho_0 \frac{1}{\omega}\left(k_x v_x+k_y v_y\right)=\rho_0\, \frac{\mu_0}{B^2_0}\,p_{\rm T}.\label{densp}
\end{align}
Density perturbations do not couple with the rest of the variables but are proportional to the total magnetic perturbation. Substituting the expressions for the velocities in the magnetic perturbation equation leads to
\begin{equation}\label{disper}
    \omega=\sqrt{k_x^2+k_y^2}\, v_{\rm A}= k\, v_{\rm A}.
\end{equation}
This dispersion relation corresponds to an isotropic fast MHD wave that propagates purely perpendicular to the magnetic field (in our case at the fast speed $v_{\rm A}$ since the sound speed is assumed to be zero). 
This MHD wave produces changes in the $x-y$ plane in velocity, magnetic field and density. 


\begin{figure}[ht!]
\centering\includegraphics[width=0.9\linewidth]{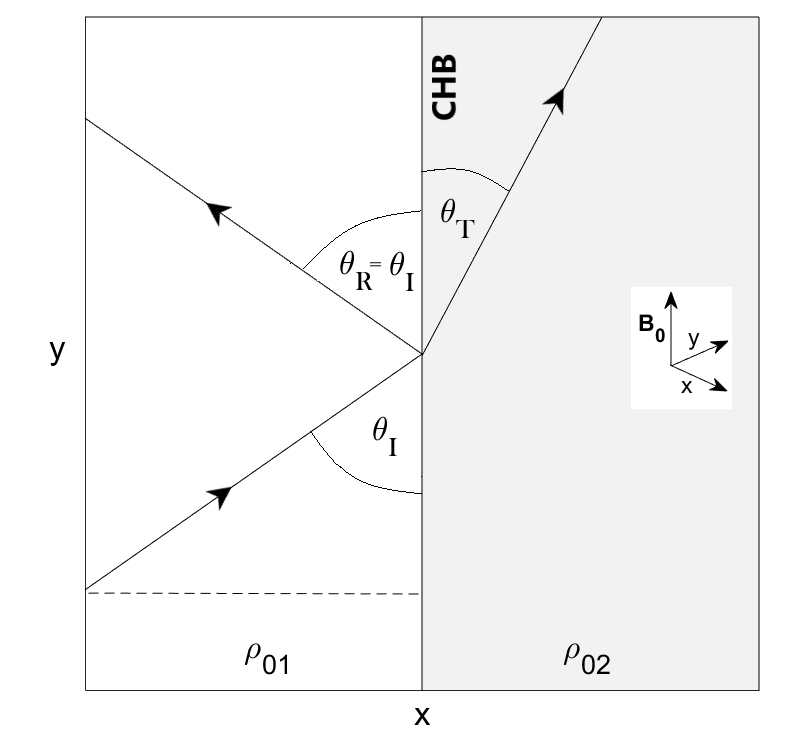}
\caption{\small Sketch of the equilibrium configuration, the two different density regions, $\rho_{01}$ and $\rho_{02}$, which are separated by the coronal hole boundary (CHB) and the wavefronts of the incident, the reflected and the transmitted wave including the angle of the incoming wave, $\theta_{\rm I}$, the reflection angle, $\theta_{\rm R}$, and the angle of the transmitted wave, $\theta_{\rm T}$. The homogeneous magnetic field, $\bf{B_0}$, points into the $z-$direction. For simplicity the density interface is located at $x=0$.}\label{sketch_incoming}
\end{figure}

\subsection{Inhomogeneous medium}

Now, that we have considered the situation for an homogeneous medium, we assume an interface in the $x-$direction which separates two regions of different densities, $\rho_{02}$ and $\rho_{01}$, representing the interior of the CH and its surrounding, see Fig.~\ref{sketch_incoming}. We define the density contrast of the two different areas as $\rho_{\rm c}=\rho_{02}/\rho_{01}$. This is an idealised representation of a CH but allows us to understand the basic properties of the reflection and the transmission in the case of a fast MHD wave at a sharp density drop. In order to understand the interaction and the evolution of a whole collection of reflected waves it is crucial to first understand single reflections and transmissions of CWs depending on the incidence angle of the incoming wave.
This situation of an oblique incident wave front at a density step has been studied in the past by, for example, \citet{stein1971} in a more general case, and in the insightful work on MHD waves of   \citet{walker2004} (see Chapter 10). We follow the latter of these approaches.

We calculate first the wavenumbers to obtain the reflection and the transmission coefficients. We assume a wave vector of the incident wave that has a component in the $y-$direction, $k_y$, parallel to the interface, and a component in the $x-$direction, $k_x$, perpendicular to the interface. Therefore, the incident wave is on the $xy-$plane. We define $\theta_{\rm I}$ as the incoming angle that lies between the incident wave vector and the interface and is restricted to be located within the range of $0\le \theta_{\rm I}\le 90^{\circ}$. We define $\theta_{\rm R}$ as the reflected angle ($\theta_{\rm R}=\theta_{\rm I}$), and $\theta_{\rm T}$ as the transmitted angle, see Fig.~\ref{sketch_incoming}. This figure illustrates that the wavenumbers in the first medium are
\begin{equation}\label{kzdefs}
    k_{y1}=k_1 \cos \theta_{\rm I},
\end{equation}
and
\begin{equation}\label{kxdefs}
    k_{x1}=k_1 \sin \theta_{\rm I}.
\end{equation}
When the incoming wave interacts with the interface the wavevector in the $y-$direction remains the same, whereas the wavenumber of the incoming wave perpendicular to the interface, $k_{x1}$, changes into the wavenumber $k_{x2}$  according to the properties of the second medium. Hence, the corresponding dispersion relation for purely fast waves in the first (homogeneous) medium is (see Eq.~(\ref{disper})) 
\begin{equation}\label{omega1}
    \omega=k_1\, v_{\rm A01}=\sqrt{k_{y1}^2+k_{x1}^2}\,v_{\rm A01},
\end{equation}
while for the second medium we obtain
\begin{equation}\label{omega2}
    \omega=k_2\, v_{\rm A02}=\sqrt{k_{y1}^2+k_{x2}^2}\,v_{\rm A02}.
\end{equation}
\noindent Since $\omega$ is constant in the reflection/transmission problem, by combining Eqs.~(\ref{kzdefs}), (\ref{omega1}) and (\ref{omega2}) with the assumption in our model that $v_{\rm A02}=v_{\rm A01} \sqrt{1/\rho_{\rm c}}$ we obtain 
\begin{equation}\label{kx2}
    k_{x2}=\pm \sqrt{\frac{\omega^2}{v_{\rm A02}^2}-k_{y1}^2}=\pm k_1 \sqrt{\rho_{\rm c}-\cos^2\theta_{\rm I}}.
\end{equation}
The question arising in this context is how to adequately choose the sign in the previous expression in order to have physically meaningful solutions. The horizontal wavenumber is either real or imaginary and therefore makes it necessary to distinguish between these two cases. In the case of a real wavenumber we have to use the positive sign because it represents a wave that propagates into the right direction inside of the second medium due to the assumed dependence of the form $e^{i\left(\omega t- k_{x2} x\right)}$. If $k_{x2}$ is imaginary, meaning an exponential dependence on $x$, we have to ensure that the wave amplitude decays with $x$ instead of growing. This is only achieved by choosing the negative sign and the character of the wave is evanescent in this case. Therefore, we rewrite the horizontal wavevector in the second medium as 
\begin{equation}\label{kx2new}
k_{x2}=k_1 \,f(\theta_{\rm I},\rho_{\rm c}),
\end{equation}
where
\begin{equation}\label{ffunct}
    f(\theta_{\rm I},\rho_{\rm c}) =
    \begin{cases}
    \sqrt{\rho_{\rm c}-\cos^2\theta_{\rm I}} & {\rm if} \quad \theta_{\rm I}\ge  \cos^{-1} \left(\sqrt{\rho_{\rm c}}\right)   \\
    -i\,\sqrt{\cos^2\theta_{\rm I}-\rho_{\rm c}}       & {\rm otherwise.}
    \end{cases}
\end{equation}
We will return to the distinction between these two situations in the following sections. The angle $\cos^{-1}(\sqrt{\rho_{\rm c}})$, that differentiates these two cases, is the so called Critical angle, $\theta_{\rm C},$ whose properties are described in detail in Sect.~\ref{crit_angle}. In optics Eqs.~(\ref{kx2new}) and (\ref{ffunct}) are identical to the equation that is obtained from the continuity of the components of {\bf E} and {\bf B} parallel to the interface that separates the two media, where $\rho_{\rm c}$ and $n_i$, the refraction index, play the same role.

\begin{figure}[ht!]
\centering \includegraphics[width=0.92\linewidth]{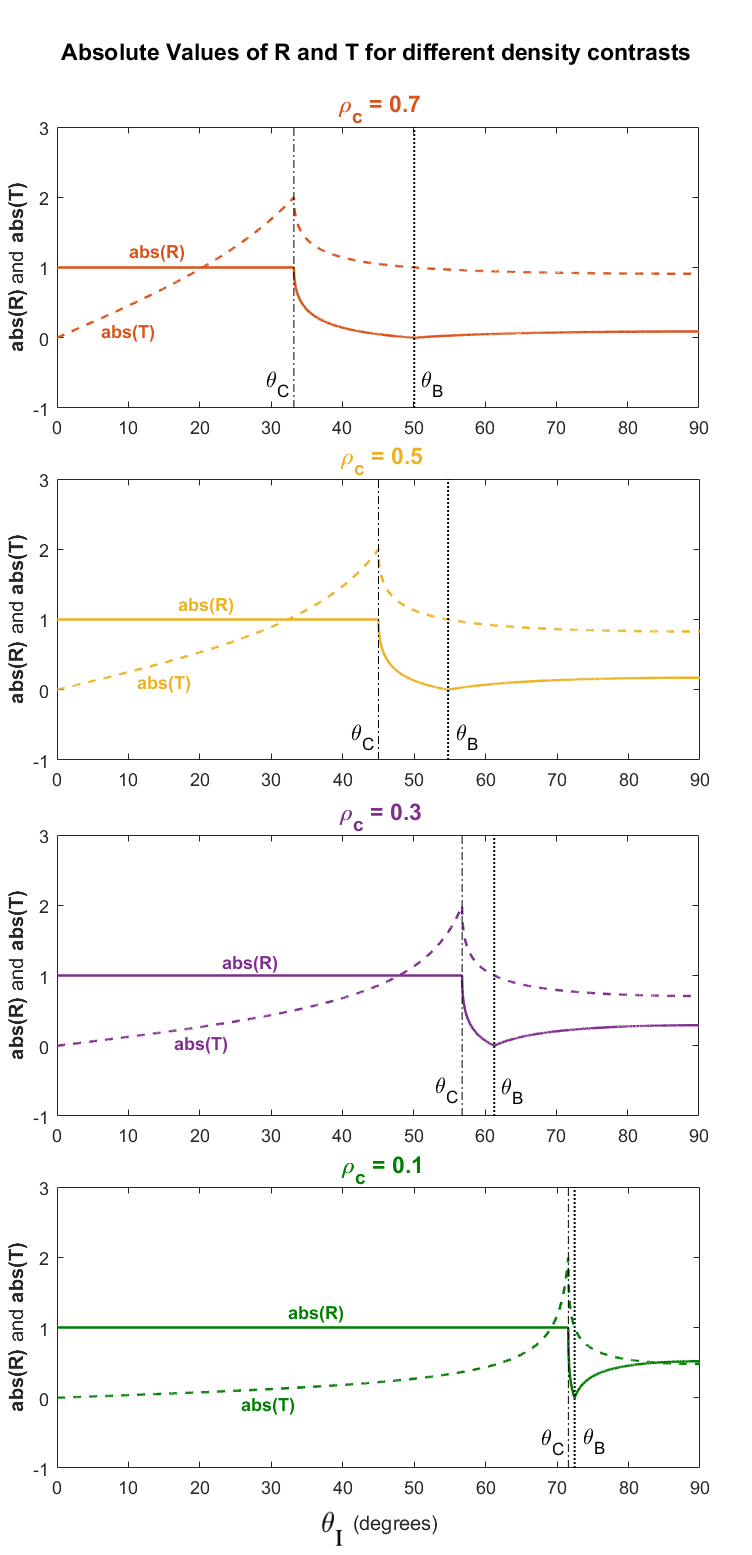}
\caption{\small Modulus of $R$ and $T$ as a function of the incidence angle, $\theta_{\rm I}$, for different values of the density contrast, $\rho_{\rm c}$. The vertical dash-dotted and dotted lines correspond to the Critical angle, $\theta_{\rm C}$, and the Brewster angle, $\theta_{\rm B}$, introduced in Sect.~\ref{angles}.}\label{abs_R_T}
\end{figure}

\begin{figure}[ht!]
\centering \includegraphics[width=\linewidth]{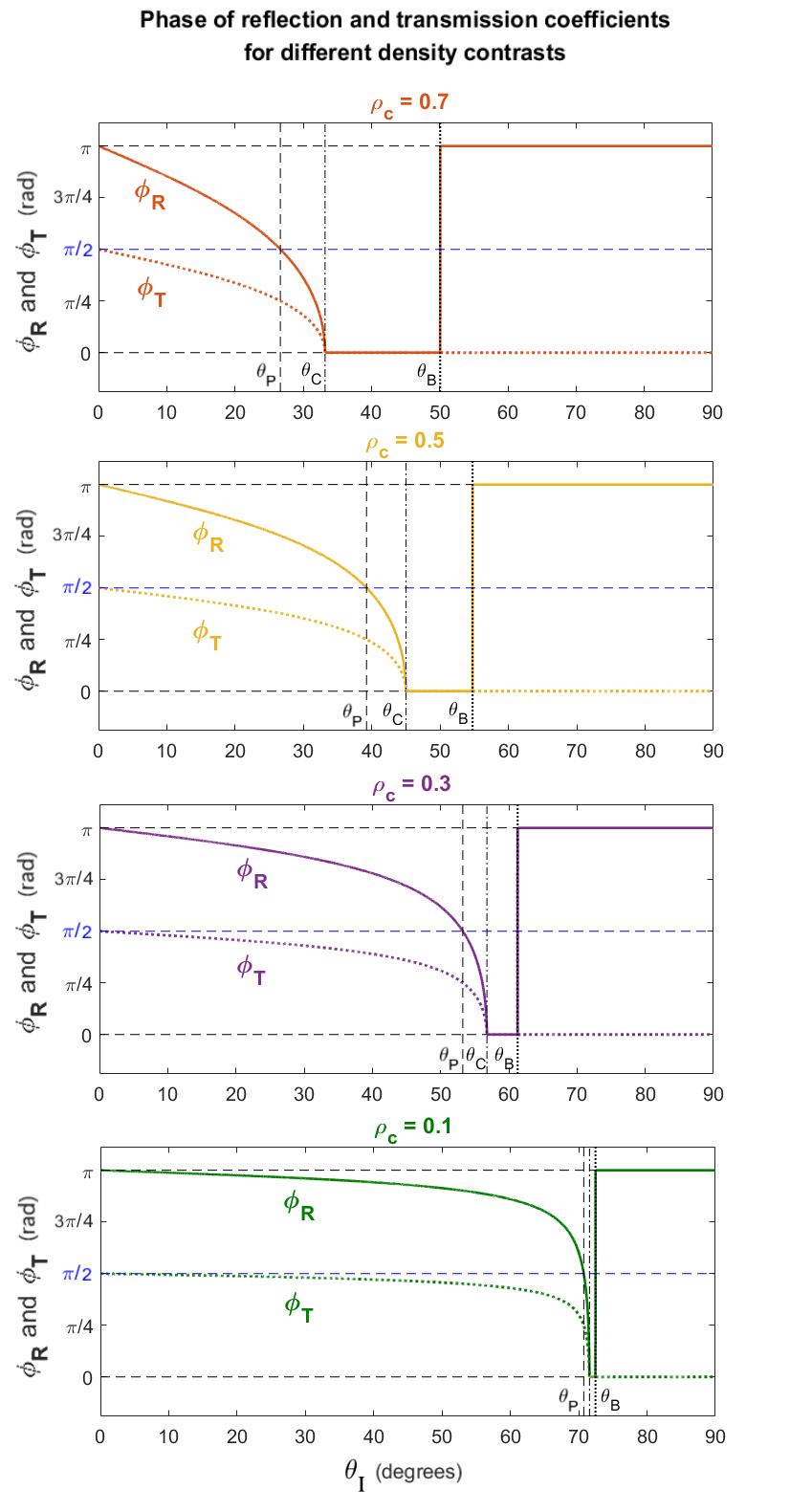}
\caption{\small Phase of $R$ and $T$ as a function of the incidence angle, $\theta_{\rm I}$, for different values of the density contrast, $\rho_{\rm c}$. The vertical dash-dotted, dotted and dashed lines correspond to the Critical angle, $\theta_{\rm C}$, the Brewster angle, $\theta_{\rm B}$, and the Phase inversion angle, $\theta_{\rm P}$, respectively, introduced in Sect.~\ref{angles}. The black dashed horizontal lines correspond to the phases $0$ and $\pi$. The blue dashed horizontal line denotes the phase difference equal to $\pi/2$ and one can see that its intersection point with $\phi_{\rm R}$ defines the Phase inversion angle, $\theta_{\rm P}$.}\label{rphasefig}
\end{figure}

Now, we focus on the conditions that the waves need to satisfy at the position of the interface, located at $x=0$ in our equilibrium model. These are the continuity of $p_{\rm T}$ and $v_x$ across the density jump. Denoting the values of these variables at each side of the interface with the superindices $+$ and $-$ according to the propagation direction of the wave and using again the subindices 1 and 2 for the first and second medium, we have that
\begin{equation}\label{contbounda}
    p^+_{\rm T1}+p^-_{\rm T1}=p^+_{\rm T2},\\
    v^+_{x1}+v^-_{x1}=v^+_{x2}.
\end{equation}
We now normalise to the magnetic pressure of the incoming wave and define the following reflection and transmission coefficients,
\begin{equation}\label{defRT}
    R=\frac{p^-_{\rm T1}}{p^+_{\rm T1}},\\
    T=\frac{p^+_{\rm T2}}{p^+_{\rm T1}}.
\end{equation}
With these definitions we have that Eq.~(\ref{contbounda}) reduces to
\begin{align}\label{condboundaeq1}
    1+R&=T,\\
    \frac{k_{x1}}{\rho_{01}\, \omega^2}\left(1-R\right)&=\frac{k_{x2}}{\rho_{02}\, \omega^2}\,T,\label{condboundaeq2}
\end{align}
where in the last equation we have used the $x-$component of the momentum equation to provide the relation between the velocity and the total perturbed magnetic pressure (see Eq.~(\ref{vx})). We also have used the fact that the reflected wave has a wavenumber that is minus the wavenumber of the incoming wave since it propagates into the negative $x-$direction. From Eqs.~(\ref{condboundaeq1}) and (\ref{condboundaeq2}) and by using the definitions for the horizontal wavenumbers it is straightforward to obtain an expression for the reflection coefficient
\begin{equation}\label{srinc}
    R(\theta_{\rm I},\rho_{\rm c})=\frac{\rho_{\rm c} \sin\theta_{\rm I} - f(\theta_{\rm I},\rho_{\rm c})}
    {
    \rho_{\rm c} \sin\theta_{\rm I} +f(\theta_{\rm I},\rho_{\rm c})},
\end{equation}
and the transmission coefficient
\begin{equation}\label{stinc}
    T(\theta_{\rm I},\rho_{\rm c})=\frac{2\,\rho_{\rm c} \sin\theta_{\rm I}}
    {
    \rho_{\rm c}\sin\theta_{\rm I} + f(\theta_{\rm I},\rho_{\rm c})}.
\end{equation}
\noindent Both, the reflection and the transmission coefficient, depend only on two parameters, the density contrast, $\rho_{\rm c}$, and the incidence angle, $\theta_{\rm I}$, and are independent of $\omega$, $k_1$ and $k_2$ in the present model. For simplicity, hereafter we use the notation $R:=R(\theta_{\rm I},\rho_{\rm c})$ and $T:=T(\theta_{\rm I},\rho_{\rm c})$. These coefficients are equivalent to Fresnel equations in optics. The coefficients derived in \citet{Piantschitsch2020} represent the special case of an incidence angle that is equal to $90^\circ$, i.e., a perpendicular incoming wave with respect to the density interface. 

The reflection and transmission coefficients are real if $f(\theta_{\rm I},\rho_{\rm c})$ is real and they are complex if $f(\theta_{\rm I},\rho_{\rm c})$ is imaginary. For this reason, we focus on the absolute values and the corresponding phases of these coefficients. The modulus of $R$ and $T$ have been plotted in Fig.~\ref{abs_R_T}. In these plots a given density ratio is assumed and the dependence on the incidence angle is shown. The modulus of $R$ is always one below the Critical angle and has an abrupt change when the behaviour of the transmitted wave changes from evanescent to propagating (for $\theta_{\rm I}=\theta_{\rm C}$). On the contrary, the modulus of $T$ changes smoothly in the two regimes but shows a sharp peak around the transition that takes place at the Critical angle. 

Since $R$ and $T$ can be complex numbers we  calculate the corresponding phases by computing the principal value of the argument. The phase, $\phi$, is represented in Fig.~\ref{rphasefig} for the same density contrasts as in Fig.~\ref{abs_R_T} and we find that it varies smoothly from a value of $\pi$, for $\theta_{\rm I}=0$, to a value of 0, for $\theta_{\rm I}=\theta_{\rm C}$. One can also see a jump from the value of 0 to $\pi$ at a the so-called Brewster angle, $\theta_{\rm B}$, which will be introduced in the following section. We will return later to this behaviour of the phase and its physical implications.

\section{Representative angles}\label{angles}


In this section we provide expressions for the transmitted angle and three relevant incidence angles which separate the area of the incoming wave into four different regions and give information about the phase and the reflection/transmission properties of secondary waves.

\subsection{Critical angle} \label{crit_angle}

As we have anticipated  in Eqs.~(\ref{kx2new}) and  (\ref{ffunct}) we can see that for certain angles the wavenumber of the second medium, which is perpendicular to the interface, $k_{x2}$, will change from real to a purely imaginary number. This implies that the wave will change its character from propagating to evanescent as it moves from the first to the second medium, meaning that there is no energy transmission but only a reflected part in the resulting waves. This happens for angles below the so-called Critical angle, defined by
\begin{equation}\label{thetc}
    \theta_{\rm C}={\rm cos^{-1}} \left(\sqrt{\rho_{\rm c}}\right).
\end{equation}
This angle is properly defined when $\rho_{02}<\rho_{01}$ (which is the case for CHs) since the argument of $\rm cos^{-1}$ must be always between $-1$ and $1$. In the case of $\rho_{01}<\rho_{02}$, there is no Critical angle, and this corresponds, for example, to any CW that propagates in the coronal medium and interacts with coronal loops which exhibit a higher density than their environment.
\begin{figure}[ht!]
\centering \includegraphics[width=0.95\linewidth]{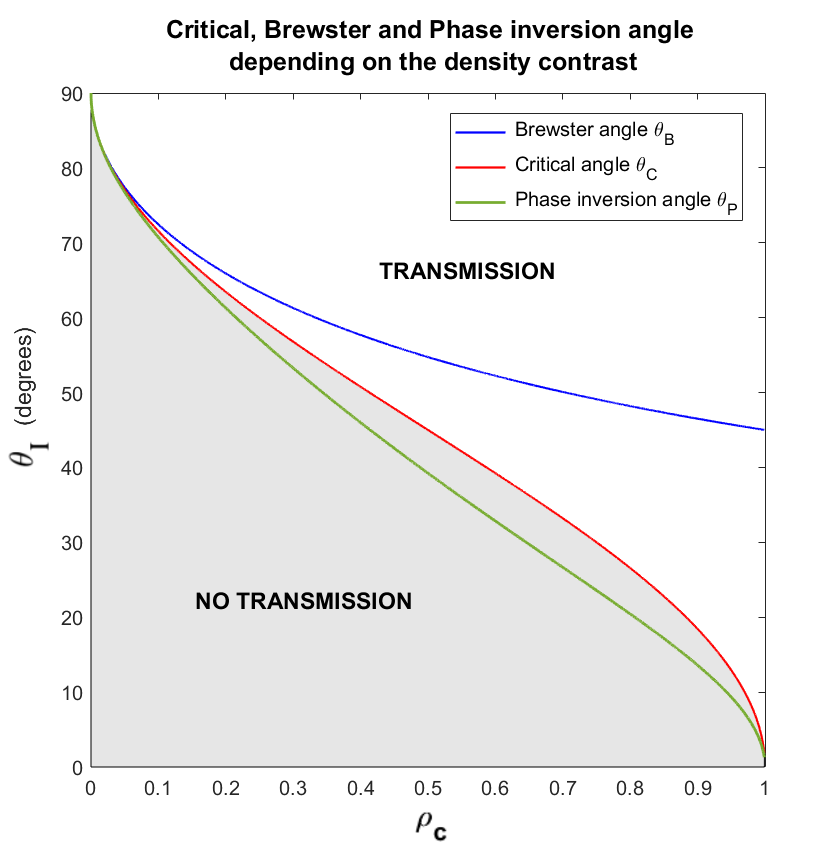}
\caption{\small Critical angle, $\theta_{\rm C}$ (Eq.~(\ref{thetc})), Brewster angle, $\theta_{\rm B}$ (Eq.~(\ref{thetb})), and Phase inversion angle, $\theta_{\rm P}$ (Eq.~(\ref{thetS})), as a function of the density contrast, $\rho_c$. Below the Critical angle (red solid line) there is no transmission of energy perpendicular to the interface (grey area), whereas for angles above the Critical there is transmission in the positive $x-$direction. The Brewster angle (blue solid line) shows the situation of perfect transmission (which is tantamount to no reflection) for different density contrasts.}\label{thetac}
\end{figure}

Fig.~\ref{thetac} corresponds to a plot of the Critical angle as a function of the density contrast. This figure clearly shows the two different regimes for the real (transmission) and the imaginary (no energy transmission) wavenumber. In the case of a homogeneous medium where $\rho_{01}=\rho_{02}$ the Critical angle $\theta_{\rm C}=0^\circ$, meaning that there is transmission for all different incidence angles which is consistent with the propagation of a wave in a single medium. If we look at the situation of a small value for the density contrast (equivalent to a large density jump when the wave enters the CH), the Critical angle, $\theta_{\rm C}$, becomes large which leads to a very restrictive transmission (see also Fig.~\ref{thetac}). For the intermediate value $\rho_{\rm c}=0.5$, the Critical angle is equal to $45^\circ$ (see Table \ref{compare_angles} and Fig.~\ref{thetac}) and separates the area of the incoming wave into two equally sized regions, of which the one containing angles smaller than $45^\circ$ leads to no transmission whereas the other part is capable of creating reflected as well as transmitted waves.

\begin{table}[ht!]
\centering
\begin{tabular}{
>{\columncolor[HTML]{FFFFFF}}c |
>{\columncolor[HTML]{FFFFFF}}c |
>{\columncolor[HTML]{FFFFFF}}c |
>{\columncolor[HTML]{FFFFFF}}c |
>{\columncolor[HTML]{FFFFFF}}c |
>{\columncolor[HTML]{FFFFFF}}c |
>{\columncolor[HTML]{FFFFFF}}c |}
\cline{2-7}
\cellcolor[HTML]{FFFFFF}                                              & \multicolumn{6}{c|}{\cellcolor[HTML]{EFEFEF}\textbf{Density contrast $\rho_c$}}                  \\ \cline{2-7} 
\multirow{-2}{*}{\cellcolor[HTML]{FFFFFF}\textbf{}}                   & \textbf{0.6} & \textbf{0.5} & \textbf{0.4} & \textbf{0.3} & \textbf{0.2} & \textbf{0.1} \\ \hline
\multicolumn{1}{|c|}{\cellcolor[HTML]{EFEFEF}\textbf{$\theta_{\rm P}$}} & 32.8           & 39.2           & 46.0           & 53.3           & 61.3           & 70.8        \\ \hline
\multicolumn{1}{|c|}{\cellcolor[HTML]{EFEFEF}\textbf{$\theta_{\rm C}$}} &    39.2          & 45.0          & 50.8          & 56.8           & 63.4          & 71.6       \\ \hline
\multicolumn{1}{|c|}{\cellcolor[HTML]{EFEFEF}\textbf{$\theta_{\rm B}$}} &   52.2           & 54.7           & 57.7           & 61.3           & 65.9           & 72.5        \\ \hline
\end{tabular}
\vspace{10px}
\caption{Critical angle, $\theta_{\rm C}$, Brewster angle, $\theta_{\rm B}$, and Phase inversion angle, $\theta_{\rm P}$ (in degrees), for different density contrasts. Observational studies derive a density ratio between $0.1$ and $0.6$ \citep[e.g.][]{Saqri2020,heinemann2019}.  }
\label{compare_angles}
\end{table}

\subsection{Brewster angle}

As already stated in Sect.~\ref{model_equat}, the reflection and the transmission coefficients, $R$ and $T$, are complex numbers when the incidence angle, $\theta_{\rm I}$, is smaller than the Critical angle, $\theta_{\rm C}$. However, in the opposite regime, $\theta_{\rm I}>\theta_{\rm C}$, it is remarkable that the numerator of the reflection coefficient is zero for the angle that satisfies
\begin{equation}\label{thetb}
    \theta_{\rm B}={\rm cos^{-1}} \left(\sqrt{\frac{\rho_{\rm c}}{1+\rho_{\rm c}}}\right).
\end{equation}
For this incidence angle all the energy of the incoming wave is transmitted into the CH, meaning that in this case we can speak about perfect transmission. This incidence angle is the equivalent to the Brewster angle defined in optics. In Table \ref{compare_angles} one can see how the value for $\theta_{\rm B}$ changes with respect to the different density contrasts. 

Using the Maclaurin series expansion for small arguments  (the typical density contrasts, $\rho_{\rm c}$, of CHs and their surrounding usually lie between $0.1$ and $0.6$) the following approximation works reasonably well 
\begin{equation}\label{thetbapp}
    \theta_{\rm B}\simeq \frac{\pi}{2}-\sqrt{\frac{\rho_{\rm c}}{1+\rho_{\rm c}}}.
\end{equation}
An important consequence of the previous equation is that the Brewster angle is always smaller than $90^\circ$ and larger than $45^\circ$ for $\rho_{\rm c} < 1$ (see Fig.~\ref{thetac}). This leads to the remarkable result that the most efficient case of energy transmission is never achieved by a CW interacting perpendicularly to the interface but always by an incidence angle that is equal to the Brewster angle, $\theta_{\rm B}$. 

Another interesting result is the fact that the smaller the value of the density contrast gets (equivalent to the density jump from outside into the CH getting larger), the closer the Critical and the Brewster angle are located to each other (see Fig.~\ref{thetac} and Table \ref{compare_angles}). This implies in particular that for CHs with a very low density compared to their surrounding only a small change of the incidence angle is sufficient to turn a case of full transmission into a case of no transmission at all. This is a crucial information for the interpretation of CW-CH interaction effects in observational data.

\subsection{Transmitted angle}

\begin{figure}[ht!]
\centering \includegraphics[width=\linewidth]{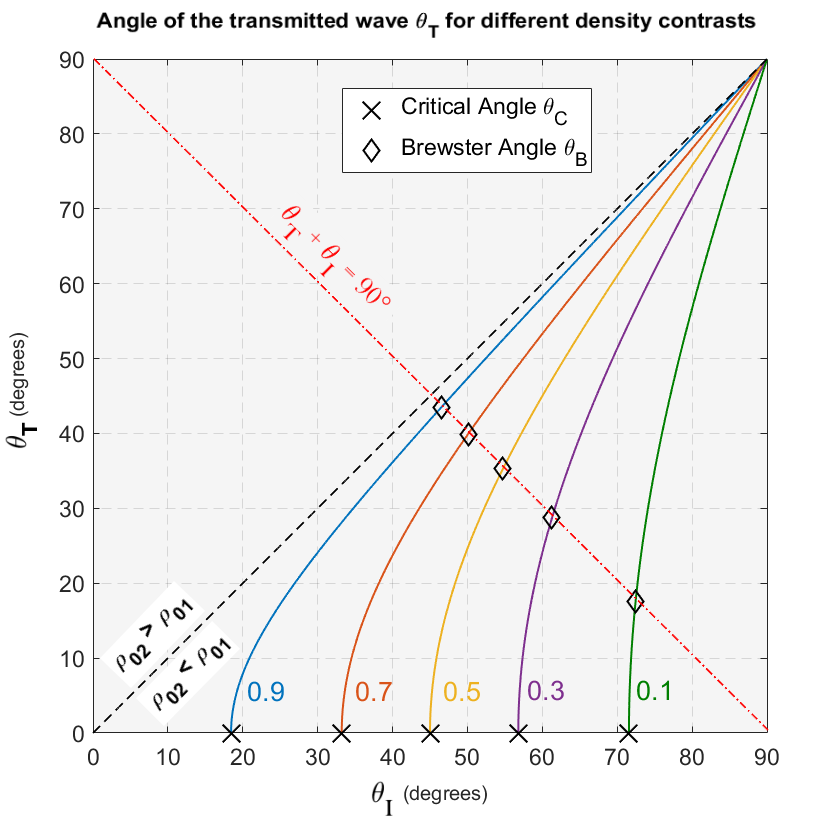}
\caption{\small Angle of the transmitted wave, $\theta_{\rm T}$ (Eq.~(\ref{thetat})), as a function of the incidence angle, $\theta_{\rm I}$, for different values of the density contrast, $\rho_c$. The intersections of the curves with the horizontal axis correspond to the Critical angle, $\theta_{\rm C}$, whereas the diamonds denote the Brewster angle, $\theta_{\rm B}$, for the different density contrasts. The red dotted line shows that $\theta_{\rm T}+\theta_{\rm B}=90^\circ$. In this figure one can see that also in the case of $\rho_{02} > \rho_{01}$ the sum of the Brewster angle and the transmitted angle is equal to $90^\circ$.}\label{thettfig1}
\end{figure}

The angle of the transmitted wave, $\theta_{\rm T}$, with respect to the incident wave can be calculated by using the information of the wavevectors (Eqs.~(\ref{kzdefs}), (\ref{kxdefs}), and (\ref{kx2new})) and applying Snell's law. This angle can be written as
\begin{equation}\label{thetat}
    \theta_{\rm T}={\rm cos^{-1}} \left(\sqrt{\frac{1}{\rho_{\rm c}}} \cos \theta_{\rm I} \right).
\end{equation}
In Fig.~\ref{thettfig1} one can see the dependence of the transmitted angle, $\theta_{\rm T}$, on the incidence angle, $\theta_{\rm I}$, for different density contrasts. Each curve, representing a different value for the density contrast, $\rho_{\rm c}$, starts at a certain threshold angle in the $x-$direction above which $\theta_{\rm T} > 0$, meaning that a transmission is possible from that value on. This threshold exactly corresponds with the Critical angle $\theta_{\rm C}$ (see marks at the $x-$axis in Fig.~\ref{thettfig1}), hence, the properties of the transmitted angle, $\theta_{\rm T}$, are consistent with those of the Critical angle, $\theta_{\rm C}$.

For angles larger than $\theta_{\rm C}$ the steepest curves are obtained for low density contrasts, implying that in these cases a smaller range of incidence angles gets transmitted which is again consistent with the larger values for $\theta_{\rm C}$ in this case.


If the incidence angle, $\theta_{\rm I}$, is close to $90^\circ$ the transmitted angle, $\theta_{\rm T}$, is close to $90^\circ$ as well, which is to be expected since in that case we are approaching the fully perpendicular case. We also find a remarkable relation between $\theta_{\rm T}$ and $\theta_{\rm B}$ in Fig.~\ref{thettfig1} where we can see the transmitted angle considered as a function of the Brewster angle for different density contrasts (see diamonds in Fig.~\ref{thettfig1}). It turns out that the sum of $\theta_{\rm T}$ and $\theta_{\rm B}$ is always equal to $90^\circ$, in the case of $\rho_{01} > \rho_{02}$ (which applies to the situation of CW-CH interaction), as well as in the opposite case  ($\rho_{01} < \rho_{02}$),  see the dotted red line in Fig.~\ref{thettfig1}. This relationship between $\theta_{\rm T}$ and $\theta_{\rm B}$ is equivalent to the situation in which the reflection coefficient, $R$, is equal to zero. These correlations are known in other contexts, such as optics, too, where in some of the cases, the Brewster angle is defined by zero reflection and, in other cases, by the assumption that the sum of the Brewster angle and its transmitted angle is equal to $90^\circ$. However, the theoretical justification of the equivalence between these two definitions is not clear in the literature and for this reason the basic steps of the corresponding proof are outlined in Appendix A.

Another important consequence of Eq.~(\ref{thetat}) is the fact that by providing the incidence angle as well as the transmitted angle we are capable of calculating the density contrast,
\begin{eqnarray}\label{thetatt}
    \frac{1}{\rho_{\rm c}}= \left(\frac{\cos \theta_{\rm T}} {\cos \theta_{\rm I}} \right)^2.
\end{eqnarray}
Measurements of the incidence and the transmitted angles are considerably easier to provide than the mean phase speed of incoming, reflected and transmitted waves which we need to get information about the density contrast in the perpendicular case \citep{Piantschitsch2020}. In general, information about the density distribution inside of a CH is mostly unavailable which makes this result a valuable supplementary tool to obtain values for density contrasts from observational data.

\subsection{Phase inversion angle}
We return now to Fig.~\ref{rphasefig} where the phases of the reflected and the transmitted waves, based on the coefficients $R$ and $T$, are plotted as a function of the incidence angle for different density contrasts. We focus first on the phase difference equal to $\pi/2$ which separates two different areas of reflected waves (see Fig.~\ref{rphasefig}) and corresponds to an angle which we call from now on the Phase inversion angle, $\theta_{\rm P}$. The real part of $R$ is zero for this angle and using Eq.~(\ref{srinc}) it is not difficult to find that
\begin{equation}\label{thetS}
    \theta_{\rm P}={\rm cos^{-1}} \left(\sqrt{\frac{\rho_{\rm c}+\rho^2_{\rm c}}{1+\rho^2_{\rm c}}}\right).
\end{equation}
In Fig.~\ref{rphasefig} one can also see that the area between the Phase inversion angle, $\theta_{\rm P}$, and the Critical angle, $\theta_{\rm C}$ gets smaller as the value of the density contrast decreases. On the contrary, large density contrasts (which correspond to values of around $0.6$ in the observations) can lead to a considerable fraction of the reflected waves which exhibit a phase difference smaller than $\pi/2$ with respect to the incoming wave. The meaning of these phase differences will be discussed in detail in Sect.~\ref{refl_wave}.

If we combine the expressions for the Critical angle (see  Eq.~(\ref{thetc})), the Brewster angle (see Eq.~(\ref{thetb})) and the Phase inversion angle (see  Eq.~(\ref{thetS})) we obtain the following relation in which all three representative angles are directly related to one another and to the density contrast,
\begin{equation}\label{relation_angles}
    \frac{\cos^2\theta_{\rm P}-{\sec^2\theta_{C}}}{\cos^2\theta_{\rm B}+{\sec^2\theta_{C}}}=\frac{\rho_{\rm c}^2-1}{\rho_{\rm c}^2+1}.
\end{equation}

At this point it is worth to mention that in Fig.~\ref{thetac} one can see that the Brewster angle, $\theta_{\rm B}$, converges to the numerical value of 45$^\circ$ while the Critical angle, $\theta_{\rm C}$, and the Phase inversion angle, $\theta_{\rm P}$, converge to 0$^\circ$ for $\rho_{\rm c}$ tending to $1$. These numerical values are also obtained by directly inserting $\rho_{\rm c}=1$ into the corresponding Eqs.~(\ref{thetb}), (\ref{thetc}) and (\ref{thetS}). However, it is important to mention that these equivalent cases are special regarding the reflection and transmission properties. A density ratio of $\rho_{\rm c}=1$ means that there is only one single medium in which the wave propagates, hence, every angle is an angle of perfect transmission and there is transmission in any case of an incoming wave which is consistent to $\theta_{\rm C}=0^\circ$.


\section{Wave energy flux}\label{wave_flux}

A physically relevant magnitude in any process involving oscillatory phenomena is the wave energy flux. In our case the energy flux has a component perpendicular to the interface but also a parallel component since we consider an oblique propagation with respect to the CH. The energy fluxes are closely related to the reflection and the transmission coefficients, $R$ and $T$, defined in Sect.~\ref{model_equat} and are capable of providing information that is needed to understand the behaviour of the secondary waves in a more comprehensive way.

In our particular situation, the wave energy flux averaged over one period does not contain the contribution of the gas pressure (since it is neglected here) and, in terms of the Poynting vector, reduces to \citep[see for example][]{walker2004}
\begin{eqnarray}\label{eflux}
\langle{\bf \Pi}\rangle=\frac{1}{2} \Re\left[\frac{\bf{E}\times{\bf b^*}}{\mu_0}\right],
\end{eqnarray}
where 
\begin{eqnarray}\label{Efield}
\bf{E}=-{\bf v}\times {\bf B_0},
\end{eqnarray}
is the perturbed electric field and $\bf b$ is the perturbation of the equilibrium magnetic field ${\bf B_0}$ introduced in Sect.~\ref{model_equat}. In Eq.~(\ref{eflux}) $\Re[z]$ denotes the real part of the complex number $z$ and the symbol * stands for the complex conjugate. By writing the velocity in terms of the magnetic pressure (Eq.~(\ref{vx})) it is not difficult to find that in our configuration the energy flux in the $x-$direction is 
\begin{eqnarray}\label{efluxx}
\langle{\Pi_x}\rangle=\frac{1}{2} \Re\left[\frac{k_x}{\rho_0\, \omega} p_{\rm T} \, p_{\rm T}^* \right],
\end{eqnarray}
and analogously, the energy flux parallel to the boundary in the $y-$direction is
\begin{eqnarray}\label{efluxy}
\langle{\Pi_y}\rangle=\frac{1}{2} \Re\left[\frac{k_y}{\rho_0\, \omega} p_{\rm T} \, p_{\rm T}^* \right].
\end{eqnarray}
These energy fluxes depend on the position and the properties of the wave (wavenumber, frequency and total pressure perturbation). It is convenient to normalise these fluxes to the incoming energy wave flux. If we do so it is not difficult to find that the normalised energy fluxes of the reflected and the transmitted waves perpendicular to the CHB (in the $x-$direction) are
\begin{equation}\label{Rdef}
    {\rm R_{ex}}=R R^*=\left|R\right|^2,
\end{equation}
and 
\begin{equation}\label{Tdef}
    {\rm T_{ex}}=\Re\left[\frac{f(\theta_{\rm I},\rho_{\rm c})}{\sin\theta_{\rm I}}\frac{1}{\rho_{\rm c}}\right]  T T^*=\frac{1}{\rho_{\rm c} \sin\theta_{\rm I}}\Re \left[f(\theta_{\rm I},\rho_{\rm c})\right]\left|T\right|^2.
\end{equation}
These expressions are similar to the reflectivity and transmissivity  coefficients defined in optics. 
Interestingly, these coefficients satisfy energy conservation (see Appendix B for a brief proof), namely
\begin{equation}\label{Econserv}
    {\rm R_{ex}}+{\rm T_{ex}}=1.
\end{equation}
Hence, in the $x-$direction the sum of the reflected wave energy flux in the first medium and the transmitted energy flux in the second medium is always equal to the incoming energy flux in the first medium. This is conceptually different to the equation $1+R=T$ for the reflection and the transmission coefficient. Indeed,  $\rm R_{ex}$ and $\rm T_{ex}$ are always real coefficients providing additional information which is difficult to obtain from the complex reflection and transmission coefficients, $R$ and $T$. 
Another useful equation can be obtained for the imaginary parts of $R$ and $T$. From Eq.~(\ref{condboundaeq1}) we can conclude that $\Im [R]=\Im [T]$ and it is not difficult to show (see Appendix B) that
\begin{equation}\label{imrnn}
   \Im[R]=\frac{1}{2\,\rho_{\rm c}\, \sin\theta_{\rm I}}\Im\left[f(\theta_{\rm I},\rho_{\rm c})^*\right]  \left|T\right|^2.
\end{equation}
This equation shows again that the coefficients $R$ and $T$ are complex if $f(\theta_{\rm I},\rho_{\rm c})$ is imaginary, otherwise they are real.

\begin{figure}[ht!]
\centering \includegraphics[width=\linewidth]{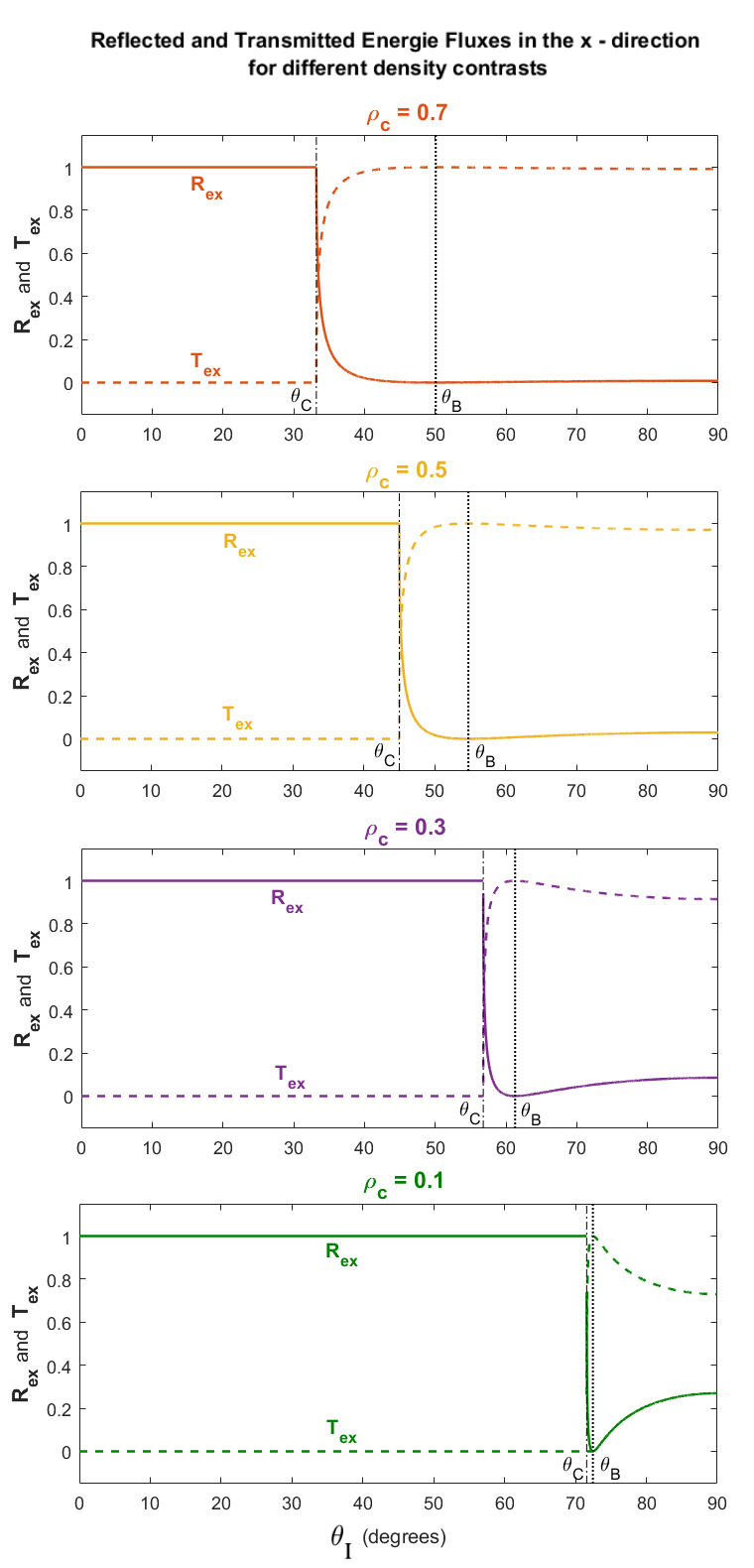}
\caption{\small Reflected (continuous line) and transmitted (dashed line) flux energies perpendicular to the interface as a function of the incidence angle, $\theta_{\rm I}$. The vertical dash-dotted and dotted lines correspond to the Critical angle, $\theta_{\rm C}$ (Eq.~(\ref{thetc})), and the Brewster angle, $\theta_{\rm B}$ (Eq.~(\ref{thetb})), respectively.}\label{retefig}
\end{figure}

The wave energy fluxes in the $x-$direction are plotted in Fig.~\ref{retefig} as a function of the incidence angle and for different density contrasts. For angles below the $\theta_{\rm C}$  there is no energy flux transmitted perpendicular to the interface (the term in the parenthesis of Eq.~(\ref{Tdef}) is an imaginary number and therefore its real part is zero). One can also see that the maximum of energy transmission happens at the Brewster angle and that the closer this angle gets to the Critical angle the smaller the value for the density contrast. Another noteworthy result that can be extracted from Fig.~\ref{retefig} is the fact that for small density contrast values the transmission coefficient for $\theta_{\rm I}\approx90^{\circ}$ decreases, meaning that a perpendicular incoming wave transmits more energy flux into the CH the larger the value of the density contrast gets, an expected result based on physical grounds.

We focus now on the energy flux parallel to the interface and we again normalise to the incoming energy flux in the first medium but in the $y-$direction. In this medium there is also a contribution, apart from the incident wave, due to the reflected wave at the boundary. The normalised reflected energy flux is simply 
\begin{equation}\label{Rdefy}
    {\rm R_{ey}}=\left|R\right|^2,
\end{equation}
which is equal to ${\rm R_{ex}}$. However, the normalised energy flux in the second medium is found to be
\begin{equation}\label{Tdefy}
    {\rm T_{ey}}=\frac{1}{\rho_{\rm c}}\left|T\right|^2,
\end{equation}
which is independent of the wavenumber parallel to the interface because the incoming and the transmitted wave have the same wavenumber $k_y$ (and we have normalised to the incoming wave flux in the $y-$direction). There is no energy conservation in the $y-$direction, since ${\rm R_{ey}}+{\rm T_{ey}}\neq 1$. The behaviour of ${\rm R_{ey}}$ and ${\rm T_{ey}}$ with the incidence angle can be inferred from Fig.~\ref{abs_R_T}.

When the wave propagates in the $x-$direction in the second medium the corresponding parallel energy flux is constant with position and given by Eq.~(\ref{Tdefy}). However, when the wave is evanescent in  the $x-$direction in this medium we have to use the proper definition given in Eq.~(\ref{kx2}) to have the energy bounded for $x$ tending to infinity. The parallel energy flux in this situation is 
\begin{equation}\label{Tey}
{\rm T_{ey}}=\frac{1}{\rho_{\rm c}}\left|T\right|^2\, e^{-2 k_1 \sqrt{\cos^2 \theta_{\rm I}-\rho_{\rm c}}\,x},
\end{equation}
and therefore the wave energy parallel to the boundary decreases exponentially with $x$ in the second medium. This allows us to define a penetration length
\begin{equation}
\lambda_P=1/(2\,k_1 \sqrt{\cos^2 \theta_{\rm I}-\rho_{\rm c}}), 
\end{equation}
which is valid only if $\theta_{\rm I}<\theta_{\rm C}$, i.e, in the evanescent regime.

\subsection{Wave energy density}

Let us briefly address the calculation of the wave energy densities. The averaged value over  a period is given by 
\begin{equation}\label{meanenerg}
    \langle{U}\rangle=\frac{1}{2} \left\{\frac{1}{2}\rho_0 \left(v_x v^*_x+v_y v^*_y\right)+\frac{1}{2}\frac{b_z b^*_z}{\mu_0}\right\},
\end{equation}
where the first term represents kinetic energy and the second term magnetic wave energy. It is not difficult to write the expressions in terms of the total pressure perturbation (using Eqs.~(\ref{vx}) and (\ref{vy}) and the definition of $p_{\rm T}$). We distinguish two situations. For incident angles above the Critical angle we have that in any of the two media the kinetic energy is
\begin{align}\label{kinener}
    \frac{1}{2}\rho_0 \left(v_x v^*_x+v_y v^*_y\right)&=\frac{1}{2}\frac{k_x^2}{\rho_0\, \omega^2}\,p_{\rm T}\, p^*_{\rm T}+\frac{1}{2}\frac{k_y^2}{\rho_0 \,\omega^2}\,p_{\rm T}\, p^*_{\rm T}=\nonumber\\
    &\frac{1}{2}\frac{\mu_0}{B^2_0}\,p_{\rm T}\, p^*_{\rm T}\left(\frac{k_x^2}{k_x^2+k_y^2}+\frac{k_y^2}{k_x^2+k_y^2}\right)=\nonumber\\
    &\frac{1}{2}\frac{\mu_0}{B^2_0}\,p_{\rm T}\, p^*_{\rm T},
\end{align}
and the magnetic energy is
\begin{align} \label{magener}  
   \frac{1}{2}\frac{b_z b^*_z}{\mu_0}&=\frac{1}{2}\frac{\mu_0}{B^2_0}\,p_{\rm T}\, p^*_{\rm T}.
\end{align}
Therefore kinetic and magnetic energies are equal meaning that there is energy equipartition in this situation. Normalising with respect to the density energy of the incident wave and using the definitions for the reflection and the transmission coefficients we obtain the simple expressions
\begin{equation}\label{energratios}
    \frac{\langle{U}\rangle_{\rm R}}{\langle{U}\rangle_{\rm I}}=|R|^2, \\  \frac{\langle{U}\rangle_{\rm T}}{\langle{U}\rangle_{\rm I}}=|T|^2.
\end{equation}
Hence, the ratios of wave energy densities are proportional to the square of the modulus of the reflection and transmission coefficients. These energy ratios are independent of the position.

Conversely, for angles below the Critical angle we have that in the first medium Eqs.~(\ref{kinener}) and (\ref{magener}) are still valid because the character of the wave is propagating, while in the second medium due to the evanescent behaviour we have now
\begin{align}\label{kinmagnevan}
    \frac{1}{2}\rho_{0} &\left(v_x v^*_x+v_y v^*_y\right)=\nonumber \\ &\frac{1}{2}\,p^+_{\rm T2} \left({p^+_{\rm T2}}\right)^* \frac{1}{\rho_{02}}\frac{|k_{x2}|^2+k^2_y}{\omega^2}\,e^{-2 k_1 \sqrt{\cos^2 \theta_{\rm I}-\rho_{\rm c}}\,x},
\end{align}
\begin{align}\label{kinmagnevan2}
    \frac{1}{2}\frac{b_z b^*_z}{\mu_0}=\frac{1}{2}\frac{\mu_0}{B^2_0} \,p^+_{\rm T2} \left({p^+_{\rm T2}}\right)^* \, e^{-2 k_1 \sqrt{\cos^2 \theta_{\rm I}-\rho_{\rm c}}\,x}.
\end{align}
There is no energy equipartition for $\theta_{\rm I}< \theta_{\rm C}$ and the energy density of the transmitted wave depends on the position. Defining again the normalised energy density with respect to the incident wave energy is not difficult to obtain that 
\begin{equation}\label{energratios1}
    \frac{\langle{U}\rangle_{\rm T}}{\langle{U}\rangle_{\rm I}}=\frac{1}{2}|T|^2\frac{1}{\rho_{02}}\frac{k^2_y}{\omega^2}\,e^{-2 k_1 \sqrt{\cos^2 \theta_{\rm I}-\rho_{\rm c}}\,x},
\end{equation}
meaning that the wave energy of the transmitted wave decreases exponentially with the distance from the location of the interface ($x=0$). It is easy to check that the above calculated magnitudes satisfy the expected fundamental relationship for plane monochromatic waves $\omega\, \langle{U}\rangle={\bf k} \cdot \langle{\bf \Pi}\rangle$ in both medium 1 and medium 2. 

The fact that there is no wave energy flux in the $x-$direction in the second medium for angles below the Critical angle does not mean that there is no wave energy in medium 2. The present wave energy calculation indeed shows that the wave energy is different from zero in medium 2 and that it decays exponentially with distance from the interface (Eq.~(\ref{energratios1})). In addition, even though in this medium ${\rm T_{ex}}= 0$, there is energy flux in the $y-$direction since ${\rm T_{ey}}\neq 0$ (Eq.~(\ref{Tey})).

\section{Reflected Wave - \textbf{depletion} or enhancement?}\label{refl_wave}

In Sect.~\ref{angles} we discussed how the three representative angles (Critical angle, Brewster angle and Phase inversion angle) separate the phase of the reflected wave into four different regions. Now, we want to analyse the physical meaning of these phase shifts with respect to the incoming wave. 

\begin{figure}[ht!]
\centering \includegraphics[width=0.9\linewidth]{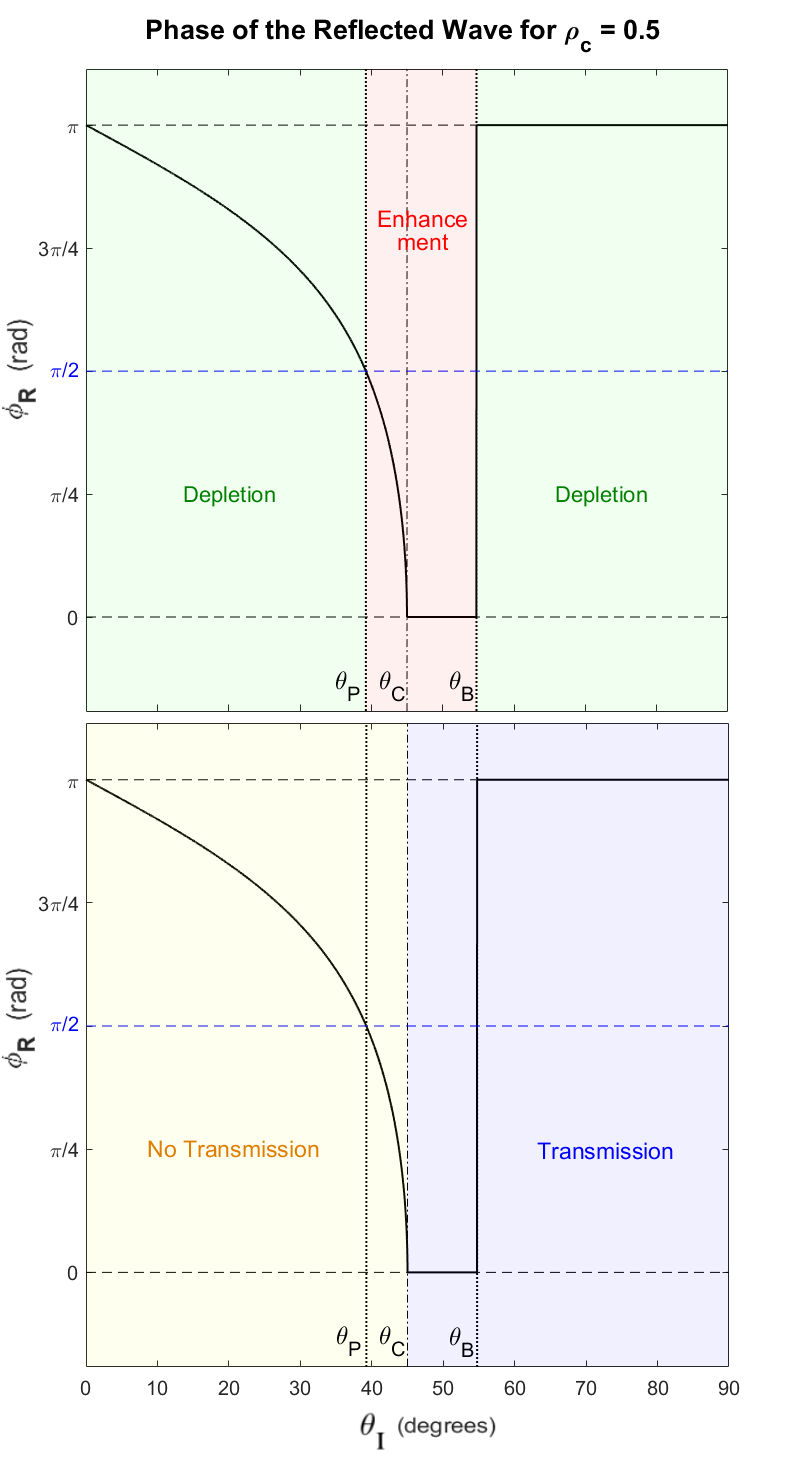}
\caption{\small {Phase of the reflected wave, $\phi_{\rm R}$, as a function of the incidence angle, $\theta_{\rm I}$. The vertical dash-dotted and dotted lines correspond to the Critical angle, $\theta_{\rm C}$, the Brewster angle, $\theta_{\rm B}$, and the Phase inversion angle, $\theta_{\rm P}$. The Phase inversion angle separates the reflected waves, which are not accompanied by a transmitted wave, into depletion and enhancement characteristics. The horizontal black dashed lines denote the phases $\pi$ and $0$. One can see that the intersection point of the horizontal blue dashed line, which represents the phase $\pi$/$2$, with the phase of the reflected wave defines the Phase inversion angle, $\theta_{\rm P}$.} }\label{phase_transm}
\end{figure}

If we assume that the incoming wave approaches the interface as a density enhancement (which is the case for CWs and CHs) then the phase differences below $\pi/2$ correspond to a reflected wave propagating as an enhancement too, whereas phase differences above $\pi/2$ imply that the reflected wave propagates as a density depletion, this is represented in Fig.~\ref{phase_transm}. In this figure and also in Fig.~\ref{rphasefig} one can see that for angles below the Critical angle, $\theta_{\rm C}$, the phase shift takes place as a smooth transition from the value $\pi$ to zero, implying that the corresponding reflected waves are either enhancements or depletions. The region furthest left in Fig.~\ref{phase_transm}, which represents phase shifts above $\pi/2$, corresponds therefore to depletions whereas the second region between $\theta_{\rm P}$ and $\theta_{\rm C}$ exhibits a smooth transition of phases from $\pi/2$ to $0$ and corresponds to an enhancement. In the third region, between $\theta_{\rm C}$ and $\theta_{\rm B}$, one can see that there is no phase difference with respect to the incoming wave which means that the corresponding reflected wave is an enhancement. The fourth region, which represents incoming angles above the Brewster angle, shows a constant phase shift of $\pi$ from the incident to the reflected wave which implies that the reflection is a density depletion. In the particular case of an incidence angle that is equal to the Brewster angle, $\theta_{\rm B}$, there is no reflection at all (see also Eq.~(\ref{thetb}) which was derived by using the assumption that the reflection coefficient $R=0$.)

By using this interpretation of reflected waves as either depletions or enhancements, we can draw another conclusion from Fig.~\ref{rphasefig}. We can see that as the value of the density contrast gets smaller, the fraction of reflected waves which display themselves as enhancements gets smaller too. On the contrary, for large density contrasts, such as $\rho_{\rm c}=0.7$, more than $20\%$ of the reflected waves propagate as enhancements since the region between $\theta_{\rm P}$ and $\theta_{\rm C}$ is larger than one fifth of the whole angle range below $\theta_{\rm C}$ in that case. This is a remarkable result since in the observations the reflected waves are usually interpreted as pure dimmings which correspond to the density depletions in our study.

However, even though the phase of the reflected waves, $\phi_{\rm R}$, tells us whether the wave propagates as an enhancement or a depletion, we still need to obtain information about the reflected density perturbation ratio in order to know how strongly the amplitude of the reflected wave increases or decreases with respect to the incoming wave. By combining Eqs.~(\ref{densp}) and (\ref{defRT}) it turns out that the density perturbation ratio can be written as  
\begin{eqnarray}\label{rho_refl_ratio}
    \frac{\rho^{-}_1}{\rho^{+}_1}=R.
\end{eqnarray}
while the velocity perturbation ratio of the reflected wave with respect to the incoming wave is simply 
\begin{eqnarray}
    \frac{v^{-}_{x1}}{v^{+}_{x1}}=-{R}.
\end{eqnarray}
We can use again Figs.~\ref{abs_R_T} and \ref{rphasefig} to understand the dependence of the reflected density and velocity perturbation on the incidence angle.

\begin{figure}[ht!]
\centering \includegraphics[width=0.95\linewidth]{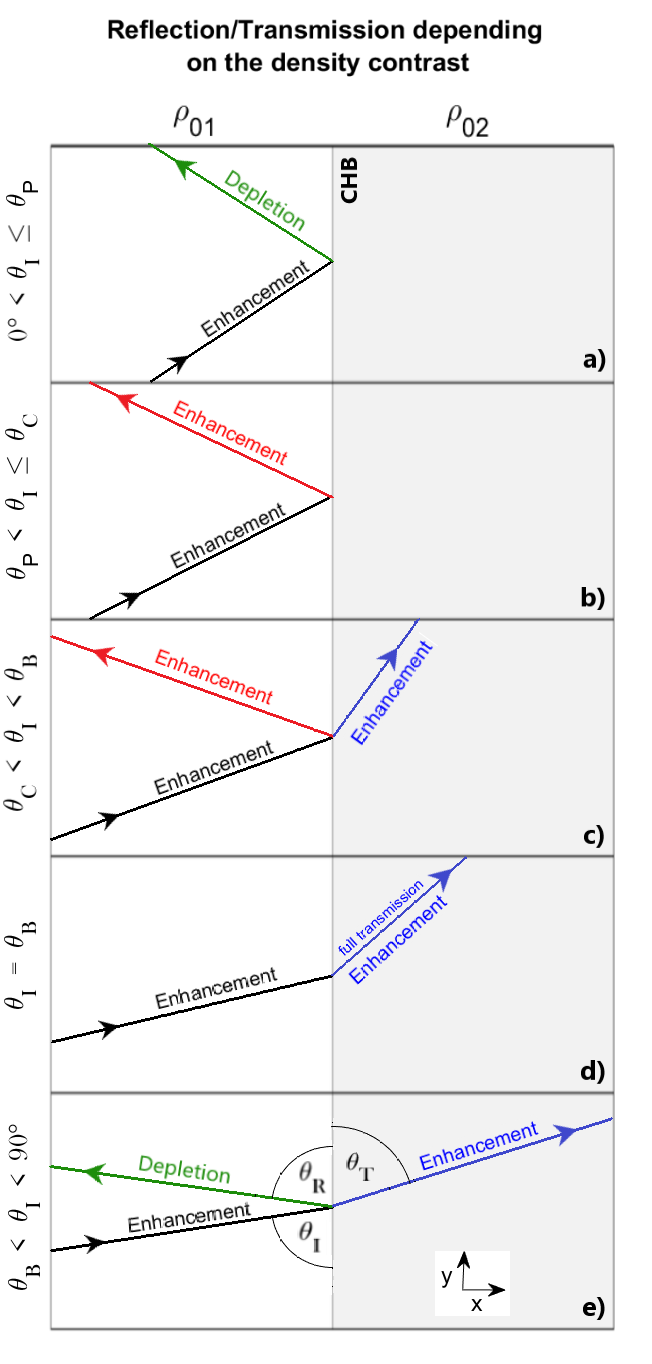}
\caption{\small Schematic representation of the interaction between an oblique incoming wave and a CHB, depending on the angle of the incident wave, $\theta_{\rm I}$. Five different scenarios are possible (assuming that the incoming wave is an enhancement): a) no transmission in the $x-$direction and the reflection is a depletion, b) no transmission and the reflection is an enhancement, c) transmission in the $x-$direction and the reflection is an enhancement, d) perfect transmission, also meaning that there is no reflection at all, e) transmission in the $x-$direction and the reflection is a depletion. }\label{angle_comp1}
\end{figure}


The lower plot in Fig.~\ref{phase_transm} complements the information of the upper plot and graphically shows in which situations the energy of the incoming wave gets transmitted into the CH (a detailed analysis of the properties of the transmitted wave is performed in the following section).

\section{Transmitted Wave - some properties}\label{transm_wave}

The three representative angles do not only separate the area of the incidence angles regarding the phase properties of the reflected wave but also regarding the behaviour of the transmitted wave. A summary of the combined results of reflection and transmission characteristics is shown in Fig.~\ref{angle_comp1} where we distinguish five different cases of CW-CH interaction (assuming the incoming wave propagates as an enhancement): a) Incidence angles within the range of $0^\circ$ and $\theta_{\rm P}$ lead to reflected waves that propagate as density depletions but do not create transmitted waves. b) The region between $\theta_{\rm P}$ and $\theta_{\rm C}$ also corresponds to no transmission but at the same time to reflected waves that propagate as enhancements. c) Incidence angles that are larger than $\theta_{\rm C}$ and smaller than $\theta_{\rm B}$ result in a situation in which there is transmission and both, reflected waves as well as transmitted waves propagate as enhancements. d) As we already stated in previous sections, if the incidence angle is equal to the Brewster angle, $\theta_{\rm B}$, there is full transmission which is equivalent to no reflection at all. e) Incidence angles that are larger than $\theta_{\rm B}$ and smaller than $90^\circ$ lead to reflected waves which propagate as depletions and transmitted waves which display themselves as enhancements. This also implies that in all cases of an actual transmission into the $x$-direction the transmitted wave propagates as an enhancement.

As in the case of the reflected wave, the density fluctuation plays an important role in understanding the behaviour of the transmission trough the interface. The density fluctuation of the transmitted wave is calculated as in the previous section by using Eq.~(\ref{densp}) which is proportional to the total pressure fluctuation. Using the definition of the transmission coefficient, $T$, which involves the total pressure fluctuation, it is easy to find that the ratio of the density perturbation associated to the transmitted wave in the second medium, $\rho^+_2$, with respect to the incident wave in the first medium, $\rho^+_1$, is
\begin{eqnarray}\label{rhopert}
    \frac{\rho^{+}_2}{\rho^{+}_1}=\rho_{\rm c} \,T.
\end{eqnarray}
The density fluctuation ratio given in Eq.~(\ref{rhopert}) is proportional to the equilibrium density contrast (which is smaller than one) but it is also proportional to the transmission coefficient that depends on the incidence angle (see Figs.~\ref{abs_R_T} and \ref{rphasefig}). We focus on the situation when there is transmission, i.e., for incidence angles that are equal or larger than $\theta_{\rm C}$, and in this case the transmission coefficient is a real number. 
\begin{figure}[ht!]
\centering \includegraphics[width=0.9\linewidth]{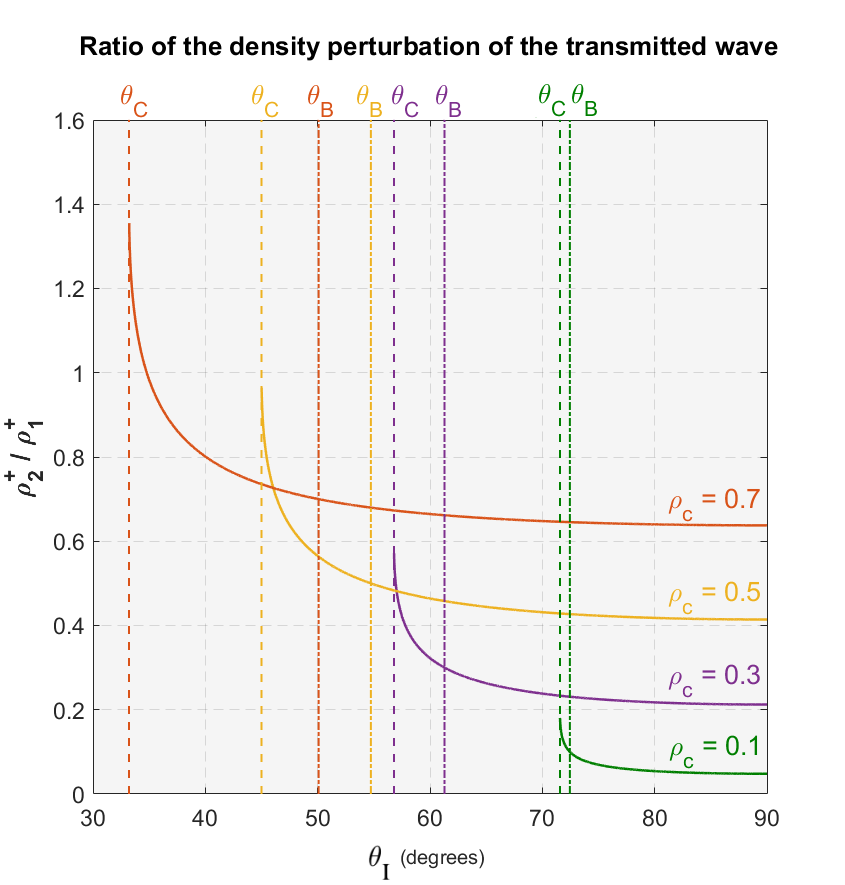}
\caption{\small Density perturbation ratio of the transmitted wave with respect to the incoming wave, $\rho^{+}_2 / \rho^{+}_1$, as a function of the incidence angle, $\theta_{\rm I}$, for different density contrasts, $\rho_c$. The vertical dashed and dashed-dotted lines denote the Critical angle, $\theta_{\rm C}$, and the Brewster angle, $\theta_{\rm B}$, for the different density contrasts.}\label{dens_pert_transm}
\end{figure}

\begin{figure}[ht!]
\centering \includegraphics[width=0.9\linewidth]{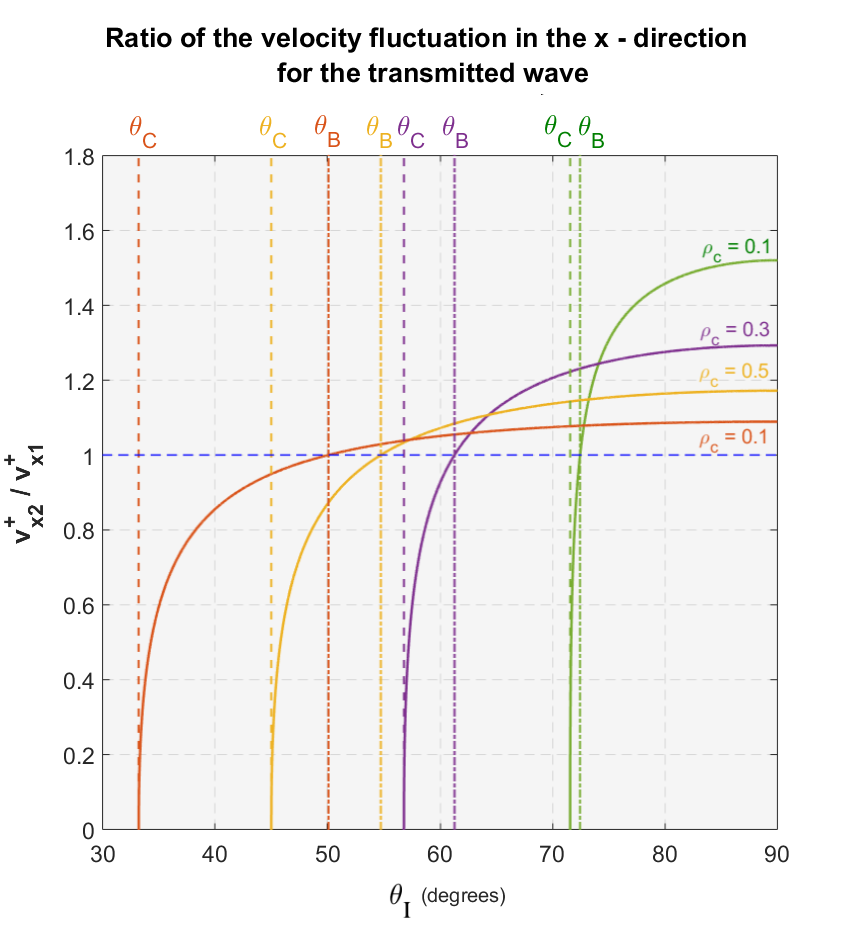}
\caption{\small Velocity perturbation ratio of the transmitted wave in the $x-$direction with respect to the incoming wave, $v^{+}_{x2} / v^{+}_{x1}$, as a function of the incidence angle, $\theta_{\rm I}$, for different density contrasts, $\rho_c$. The vertical dashed and dashed-dotted lines denote the Critical angle, $\theta_{\rm C}$, and the Brewster angle, $\theta_{\rm B}$, for the different density contrasts. The horizontal blue dashed line shows that the intersection of the velocity perturbation and the Brewster angle is equal to one for all different density contrasts. }\label{vel_pert}
\end{figure}

In Fig.~\ref{dens_pert_transm} we have plotted the density fluctuation ratio calculated using Eq.~(\ref{rhopert}) as a function of the incidence angle for different density contrasts of the equilibrium. One can see that for equilibrium density contrasts smaller than $0.5$ the ratio is always below one which means that in these cases the amplitude of the transmitted wave decreases with respect to the incoming one. For density contrasts larger than $0.5$ we find that density ratios of values larger than one are possible, more precisely, if the large density contrast is accompanied by an incidence angle that lies close to the Critical angle, $\theta_{\rm C}$. This is equivalent to an increase of the density amplitude of the transmitted wave with respect to the incoming one. This result is remarkable for two reasons: First, in the perpendicular case the density perturbation ratio is smaller than one for any density contrast. Second, density amplitudes of transmitted waves have so far been usually assumed to decrease when the CW enters the CH.   

We concentrate now on the velocities. For the propagating case the velocity component perpendicular to the interface is 
\begin{eqnarray}\label{rho_refl_ratio2}
    \frac{v^{+}_{x2}}{v^{+}_{x1}}=\frac{1}{\rho_c}\frac{\sqrt{\rho_c-\cos^2\theta_{\rm I}}}{\sin\theta_{\rm I}}\,T.
\end{eqnarray}
In Fig.~\ref{vel_pert} we have plotted the perpendicular velocity fluctuation ratio as a function of the incidence angle for different density contrasts. One can see that the intersection points of the ratio functions and the regarding Brewster angles are located exactly on the horizontal blue dashed line of value one. This implies that incidence angles that are equal to the Brewster angle lead to situations in which the velocity amplitude of the transmitted wave is exactly the same as the incoming one. Nevertheless, for angles above the Brewster angle we find that the velocity amplitude is larger than one. The reason of this behaviour is understood recalling that the total kinetic energy is exactly the same as the magnetic energy but the kinetic energy has two contributions from the $x$ and $y$ velocity components (see Eq.~(\ref{kinener})). This essentially means that when the $y$ velocity component decreases the $x$ component must increase, and this is precisely the situation when the incidence angle is increased. 

Regarding the velocity parallel to the density jump, it is worth to mention that, in contrast to $v_x$, now $v_y$ is not continuous across the CHB. The parallel velocity is given by Eq.~(\ref{vy}). It is not difficult to see that the parallel velocity on the first medium at the boundary ($x=0$) is $1+R$ (incoming plus reflected) while for the second medium we have $ T / \rho_{\rm c}$ (normalised to the incoming wave). Hence, the ratio of the parallel velocities in the second medium with respect to the first medium is simply $1/\rho_{\rm c}$ (since $1+R=T$). The inverse of the density contrast is therefore a direct measure of the jump in the velocity parallel to the boundary.

Finally, we describe how a wavefront changes as it interacts with the interface. This deserves attention since observations can provide some information although in general the waves transmitted into the CH are very faint and difficult to track. The wave model presented here assumes purely harmonic sinusoidal waves (e.g. Eq.~(\ref{simplewave})), but real CW perturbations have a front of finite width. Nevertheless, we know that we can decompose the shape of the pulse in a superposition of plane waves using Fourier analysis and use the results derived in this work. From the theoretical point of view every single wavenumber changes from the first to the second medium. Using Eqs.~(\ref{omega1}) and  (\ref{omega2}) we have that 
\begin{eqnarray}
    \frac{k_1}{k_2}=\frac{v_{\rm A02}}{v_{\rm A01}}=\sqrt{\frac{1}{\rho_{\rm c}}}.
\end{eqnarray}
Since all the wavenumbers of the pulse change in the same way, if we have an incident pulse with a typical spatial width of $\lambda_{\rm inc}$ the transmitted pulse has a width of $\lambda_{\rm trans}$, and both are related through
\begin{eqnarray}\label{lambdas}
    \frac{\lambda_{\rm trans}}{\lambda_{\rm inc}}=\sqrt{\frac{1}{\rho_{\rm c}}}.
\end{eqnarray}
For CHs we have that $\rho_{02}<\rho_{01}$ ($\rho_{\rm c}<1$) and the width of the transmitted pulse is always larger than the one of the incident wave. Hence, the density contrast can be directly inferred from the information, if available, of the widths of the incoming and transmitted pulses using Eq.~(\ref{lambdas}). Another path to determine the density contrast is given in Eq.~(\ref{thetatt}) but using the information of the incidence and transmitted angles.


\section{Discussion and Conclusions}\label{conclusions}

In this study we described and analysed the geometrical properties of CW-CH interactions including an oblique incoming CW with respect to the CHB. Using linear MHD theory and considering a fast MHD wave approaching a region of low density at any angle we found remarkable features that are relevant regarding the interpretation of the observations. The results presented in this study are based on the reflection and the transmission coefficients and allow us to comprehensively understand the details of the CW-CH interaction process.

The main theoretical results are summarised as follows:
\begin{enumerate}
\item{The analytical expressions for the reflection and the transmission coefficients (see 
Eqs.~(\ref{srinc}) and (\ref{stinc})) depend only on the incidence angle and the density contrast between the CH and its surrounding. These coefficients are used to calculate the wave energy flux, the phase differences as well as the density/velocity amplitudes of the reflected and transmitted waves with respect to the incoming wave.}

\item{The Critical angle (see Eq.~(\ref{thetc})) separates the region of the incoming wave into an area which does not allow energy transmission perpendicular to the interface and a second area that leads to energy transmission into the second medium (see Fig.~\ref{phase_transm}). It is important to remark that even in the case of no wave energy transmission in the perpendicular direction there is wave energy in the second medium and also energy flux parallel to the boundary inside the CH, both exponentially decaying from the location of the interface.}

\item{We showed that an incidence angle that is equal to the Brewster angle (see Eq.~(\ref{thetb})) implies perfect transmission which is equivalent to no reflection at all (see Fig.~\ref{angle_comp1}). Hence, a perpendicular incoming wave does not lead to the most efficient situation regarding energy transmission.}

\item{We found that the Phase inversion angle (see Eq.(\ref{thetS})), which corresponds to the phase shift $\pi/2$ of the reflected wave with respect to the incoming wave, separates the area below the Critical angle into two different regions (see Figs.~\ref{phase_transm} and  \ref{angle_comp1}). If the incidence angle is located within the range of $\theta_{\rm P} < \theta_{\rm I} <\theta_{\rm C}$, then the reflected wave propagates as an enhancement (if the incoming wave is an enhancement too). If the incidence angle is smaller than the Phase inversion angle, $\theta_{\rm P}$, the reflected wave propagates as a depletion. All the three representative angles, $\theta_{\rm C}$, $\theta_{\rm B}$ and $\theta_{\rm P}$ depend only on the density contrast in the present model. }

\item{We provided an analytical expression for the angle of the transmitted wave (see Eq.~(\ref{thetat})) which depends on the density contrast as well as the incidence angle. This equation also gives us the possibility to calculate the density contrast in case measurements of incidence angle and transmitted angle can be provided (see Eq.~(\ref{thetatt})). Since the density inside a CH is mostly unavailable, this results serves as an additional tool to estimate density contrasts of CHs and their surroundings. }

\item{We provided analytical expressions for the density and the velocity amplitudes of the reflected as well as the transmitted waves. We found that for large density contrast values in combination with incidence angles which lie close to the Critical angle, the density amplitude of the transmitted wave can be larger than the one of the incoming wave.}

\item{We found that the smaller the value of the density contrast, the closer the Critical angle and the Brewster angle lie next to each other. This implies that for CHs with a low density a small change of the incidence angle is sufficient to turn a case of full transmission into a case of no transmission at all which leads to important consequences regarding the interpretation of secondary waves in the observations.}

\item{Since our model assumes a phase speed  of $v_{\rm A02}=v_{\rm A01} \sqrt{\rho_c}$ in the second medium, we are able, by using the information in point 5, to obtain the phase speed of the transmitted wave by providing the incidence angle, the transmitted angle and the incoming phase speed.}

\item{In summary, an oblique incoming wave with respect to the density interface can lead to five different cases (assuming the incoming wave is an enhancement):
\begin{figure}[hh!]
\centering \includegraphics[width=0.9\linewidth]{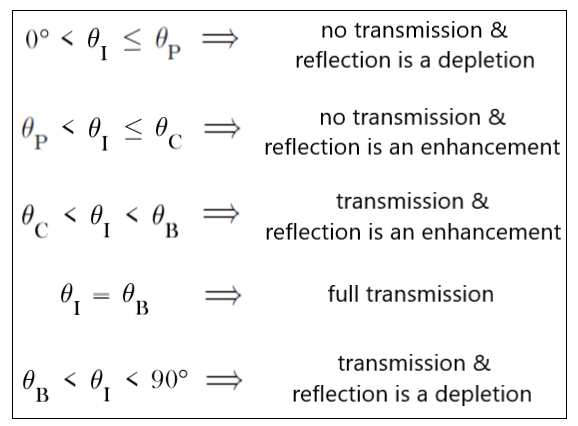}
\end{figure}
}
\end{enumerate}

\vspace{20px}
We have to keep in mind that an actual CW that interacts with a CH is an entire wave front and leads to a whole collection of reflected waves, hence, the secondary waves have to be considered as the superposition of different secondary waves. However, in order to understand this process it is crucial to understand first the properties of single reflections and transmissions which we have studied in this paper.

Several of the results obtained in this study show a surprising behaviour regarding the interpretations of secondary waves, such as the distinction between the depletions and enhancements of the reflected waves which are usually interpreted as pure dimmings, and hence as density depletions in the observations. Another remarkable result is the fact that the density amplitudes of the transmitted wave can be larger than the incoming ones in certain cases. These results might lead to a new interpretation of already studied events and have the potential to partially explain ambiguous CW parameters.

Another noteworthy result is the fact that the shape of perturbation for a reflected wave does not change due to the interaction at the interface between the two media. The reflected wave exhibits the same shape just with a different global amplitude and phase. But for the transmitted wave, travelling always at a faster speed than the incoming wave, the width of the perturbation increases by a factor $1/\sqrt{\rho_{\rm c}}$ and it is independent of the incidence angle.

Since we have used a linear theory to derive our theoretical results it is also important to mention that the intensity/density enhancements in our model correspond to shocks in the nonlinear regime while a depletion corresponds to a rarefaction, which is essential when it comes to interpretation of observational results.

The model considers an idealised case of CW-CH interaction including a simplified shape of the CH, zero gas pressure, a sharp gradient representing the interface between the two media and a homogeneous magnetic field. On the one hand, this simplification allowed us to derive analytical expressions for CW parameters which can serve as a fast and useful tool to calculate density contrasts as well as phase speeds and other properties of secondary waves. On the other hand, the comparison between these analytical expressions derived from a linear theory and the simulation results of weakly nonlinear MHD waves, described in \citet{Piantschitsch2020}, showed good agreement and therefore validate the approach followed in this study. 

In order to consider a more complex CW-CH interaction model, including more realistic parameters and whole wave fronts in 2D, one has to perform MHD simulations which will be addressed in future studies but also comparisons to observational data of CW-CH interaction events.

The results presented in this work can be visualised using a newly developed tool that the user can find at \url{https://editor.p5js.org/jaumeterradascalafell/sketches/P-T_SVKnc}. This JavaScript code allows to modify the main input parameters which are the density contrast of the CH and the incidence angle of the CW. The main output is a movie of the incoming, the reflected and the transmitted wave (if any) along with the corresponding amplitudes, wavelengths and propagation speeds.

It is worth to mention that in the Javascript code the incoming wave is assumed to have a Gaussian shape instead of a sinusoidal function. In a realistic situation there are observational indications that the incoming wave is close to a single pulse and not a periodic structure as it has been assumed in the analytical part of the present work. However, a spatially localised perturbation like a Gaussian can be interpreted as a superposition/combination of different plane harmonic waves. Due to the fact that the dispersion relation is independent of the wavenumber all the wavelengths that form part of the pulse travel at the same speed and the pulse keeps its shape constant when traveling in the same medium. Therefore, all the wavelengths contained in the pulse with show the same phase difference when they are reflected at the interface. This means that the pulse will change its amplitude (becoming negative depending on the parameters) during the reflection but the shape is maintained. This behaviour is visualised using the online tool created for this purpose.

\begin{acknowledgements} I.P. and J.T. acknowledge the support from grant AYA2017-85465-P
(MINECO/AEI/FEDER, UE), to the Conselleria d'Innovaci\'o, Recerca i Turisme del
Govern Balear, and also to IAC$^3$. This work was supported by the Austrian Science Fund (FWF): I 3955-N27. We are grateful to Marc Carbonell from the Departament de Ci\`encies Matem\`atiques i Inform\`atica, Universitat de les Illes Balears (UIB), for his advice on the derivations shown in the Appendix. We also thank the p5.js project, currently led by Moira Turner and created by Lauren Lee McCarthy. p5.js is developed by a community of collaborators, with support from the Processing Foundation and NYU ITP. 
\end{acknowledgements}
 \vspace{-30px}
\bibliographystyle{aa}      
\bibliography{paper_2_cit}   

\appendix

\section{ }
In the literature, in particular in the field of optics, the Brewster angle is defined by zero reflection ($R=0$) in some contexts. In other contexts, the assumption that the sum of the Brewster angle, $\theta_{\rm B}$ and its transmitted angle, $\theta_{\rm T}$, is equal to $90^\circ$ is used to characterise the Brewster angle. However, the theoretical justification that these definitions are equivalent to each other is not straightforward and for this reason we will use the definition of the Brewster angle in this paper (see Eq.~(\ref{thetb})) to outline the basic steps of the according proof here. 

In the following we assume that the incidence angle, $\theta_{\rm I}$ and the Brewster angle, $\theta_{\rm B}$ are above the critical angle to have transmission, while the transmitted angle, $\theta_{\rm T}$, lies between $0^\circ$ and $90^\circ$. Therefore the function $f(\theta_{\rm I},\rho_{\rm c})$ is always real (see Eq.~(\ref{ffunct})).

We start with the relation of the sum of the Brewster and transmitted angle and we want to prove that $R=0$. Therefore, 
\begin{eqnarray}
 \theta_{\rm T}+\theta_{\rm B}=90^\circ,
\end{eqnarray}
which is equivalent to,
\begin{eqnarray}
 \cos\left(\theta_{\rm T}+\theta_{\rm B}\right)=0.
\end{eqnarray}
Using the addition formula for the cosine we have,
\begin{eqnarray}
 \cos\theta_{\rm B}\,\cos\theta_{\rm T}=\sin\theta_{\rm B}\,\sin\theta_{\rm T}.
\end{eqnarray}
We use the definition of Brewster (Eq.~(\ref{thetb})) and transmitted angle (Eq.~(\ref{thetat})) on the left-hand side and the Pythagorean trigonometric identity on the right-hand side
\begin{eqnarray}
\sqrt{\frac{\rho_{\rm c}}{1+\rho_{\rm c}}}\,\sqrt{\frac{1}{\rho_{\rm c}}} \,\cos\theta_{\rm I}=
\sqrt{1-\cos^2\theta_{\rm B}}\,\sqrt{1-\cos^2\theta_{\rm T}},
\end{eqnarray}
and using again the definitions of the angles but now on the right-hand side we have
\begin{eqnarray}
\sqrt{\frac{1}{1+\rho_{\rm c}}}\, \cos\theta_{\rm I}=
\sqrt{1-\frac{\rho_{\rm c}}{1+\rho_{\rm c}}}\, \sqrt{1-\frac{1}{\rho_{\rm c}} \cos^2\theta_{\rm I}}.
\end{eqnarray}
Cancelling multiplicative constants leads to
\begin{eqnarray}
\cos\theta_{\rm I}=\sqrt{1-\frac{1}{\rho_{\rm c}} \cos^2\theta_{\rm I}},
\end{eqnarray}
and taking the square on both sides
\begin{eqnarray}
\cos^2\theta_{\rm I}=1-\frac{1}{\rho_{\rm c}} \cos^2\theta_{\rm I},
\end{eqnarray}
which is rewritten as
\begin{eqnarray}
\left(1+\frac{1}{\rho_{\rm c}}\right)\cos^2\theta_{\rm I}=1.
\end{eqnarray}
This is an important step, we multiply the previous equation by $\left(1-\frac{1}{\rho_{\rm c}}\right)$, assuming that $\rho_{\rm c} \ne 0$), we have
\begin{eqnarray}
\left(1-\frac{1}{\rho^2_{\rm c}}\right)\cos^2\theta_{\rm I}=1-\frac{1}{\rho_{\rm c}},
\end{eqnarray}
obtaining
\begin{eqnarray}
\cos^2\theta_{\rm I}-\frac{1}{\rho^2_{\rm c}}\cos^2\theta_{\rm I}=1-\frac{1}{\rho_{\rm c}},
\end{eqnarray}
which is rewritten as
\begin{eqnarray}
\frac{1}{\rho_{\rm c}}-\frac{1}{\rho^2_{\rm c}}\cos^2\theta_{\rm I}=1-\cos^2\theta_{\rm I}.
\end{eqnarray}
We use again the Pythagorean trigonometric identity on the right-hand side, 
\begin{eqnarray}
\frac{1}{\rho^2_{\rm c}}\left(\rho_{\rm c}-\cos^2\theta_{\rm I}\right)=\sin^2\theta_{\rm I},
\end{eqnarray}
and we take the square root
\begin{eqnarray}
\frac{1}{\rho_{\rm c}}\sqrt{\rho_{\rm c}-\cos^2\theta_{\rm I}}=\sin\theta_{\rm I}.
\end{eqnarray}
This last equation is rewritten as 
\begin{eqnarray}
\sin\theta_{\rm I}-\frac{1}{\rho_{\rm c}}\sqrt{\rho_{\rm c}-\cos^2\theta_{\rm I}}=0,
\end{eqnarray}
which is equivalent to (see Eq.~(\ref{srinc}))
\begin{eqnarray}
R=0.
\end{eqnarray}
This is the result we wanted to prove. The steps performed in this proof are valid in the reverse direction too and show therefore equivalence of the two assertions.

\section{ }
We briefly give the proof of energy conservation in energy flux perpendicular to the interface. We write Eqs.~(\ref{condboundaeq1}) and (\ref{condboundaeq2}) as
\begin{align}\label{condboundaeqnew}
    1+R&=T,\\
    1-R&=\frac{k_{x2}}{k_{x1}}\frac{1}{\rho_{\rm c}}\,T.\label{condboundaeqnew1}
\end{align}
The complex conjugate of (\ref{condboundaeqnew1}) is
\begin{eqnarray}\label{condboundaeqnew1n}
    1-R^*=\left(\frac{k_{x2}}{k_{x1}}\frac{1}{\rho_{\rm c}}\right)^*\,T^*.
\end{eqnarray}
We multiply (\ref{condboundaeqnew}) by (\ref{condboundaeqnew1n}) to obtain 
\begin{eqnarray}\label{condboundaeqnew1nn}
    1+R-R^*-\left|R\right|^2=\left(\frac{k_{x2}}{k_{x1}}\frac{1}{\rho_{\rm c}}\right)^*\,\left|T\right|^2.
\end{eqnarray}
The imaginary part of (\ref{condboundaeqnew1nn}) leads to the following equation
\begin{equation}\label{imr}
   \Im[R]=\frac{1}{2\,k_{x1}\, \rho_{\rm c}} \Im\left[k_{x2}^*\right]  \left|T\right|^2,
\end{equation}
while the real part is
\begin{eqnarray}\label{condboundaeqnew1nnn}
    1-\left|R\right|^2=\frac{1}{k_{x1}\, \rho_{\rm c}}\Re \left[k_{x2}\right]\,\left|T\right|^2,
\end{eqnarray}
where $\Re[z]$ is the real part of the complex number $z$ and $\Im[z]$ the imaginary part. Substituting the expressions for the horizontal wavenumbers Eqs.~(\ref{kxdefs}) and (\ref{kx2new}) into (\ref{condboundaeqnew1nnn}) we finally obtain
\begin{equation}\label{imrn}
   \Im[R]=\frac{1}{2\,  \rho_{\rm c} \, \sin\theta_{\rm I}}\Im\left[f(\theta_{\rm I},\rho_{\rm c})^*\right]\left|T\right|^2,
\end{equation}
and
\begin{equation}\label{contbf}
   1-\left|R\right|^2=\frac{1}{\rho_{\rm c}\, \sin\theta_{\rm I}}\Re\left[f(\theta_{\rm I},\rho_{\rm c})\right]  \left|T\right|^2.
\end{equation}
Using the definitions for $\rm R_{ex}$ and $\rm T_{ex}$ (Eqs.~(\ref{Rdef}) and (\ref{Tdef})) the last equation leads to the energy flux conservation in Eq.~(\ref{Econserv}),
\begin{equation}\label{Econservn}
    {\rm R_{ex}}+{\rm T_{ex}}=1.
\end{equation}

\end{document}